\input ./style/arxiv-general.cfg
\documentclass[aoas,MSNbibl,nameyear,seceqn,dvips]{arximspdf}
\makeatletter
   \@ifpackageloaded{graphicx}{}{\usepackage{graphicx}}
\makeatother
\usepackage{url,breakurl}

\doi{10.1214/15-AOAS861}
\volume{9}
\issue{4}
\pubyear{2015}
\firstpage{1726}
\lastpage{1760}
\docsubty{FLA}

\makeatletter
\def\epsilon{\varepsilon}

\newcommand{\eqref}[1]{(\ref{#1})}
\makeatother

\begin{document}
\begin{frontmatter}

\title{The discriminative functional mixture model for a~comparative
analysis of bike sharing systems}
\runtitle{The DFM model for a comparative analysis of BSS}

\begin{aug}
\author[A]{\fnms{Charles}~\snm{Bouveyron}\corref{}\ead[label=e1]{charles.bouveyron@parisdescartes.fr}},
\author[B]{\fnms{Etienne}~\snm{C\^ome}\ead[label=e2]{etienne.come@ifsttar.fr}}
\and
\author[C]{\fnms{Julien}~\snm{Jacques}\ead[label=e3]{julien.jacques@univ-lyon2.fr}}

\runauthor{C. Bouveyron, E. C\^ome and J. Jacques}

\affiliation{Universit\'e Paris Descartes,
IFSTTAR and
Universit\'e Lumi\'ere Lyon 2}

\address[A]{C. Bouveyron\\
Laboratoire MAP5\\
Universit\'e Paris Descartes\\
45 rue des Saints P\`{e}res\\
75006 Paris\\
France\\
\printead{e1}}
\address[B]{E. C\^ome\\
IFSTTAR, Grettia\\
14-20 Boulevard Newton\\
Cit\'{e} Descartes, Champs sur Marne\\
77447 Marne la Vall\'{e}e Cedex 2\\
France\\
\printead{e2}}
\address[C]{J. Jacques\\
Laboratoire ERIC\\
Universit\'e Lumi\'ere Lyon 2\\
Bureau K067\\
5 av. Mend\`{e}s France\\
69676 BRON Cedex\\
France\\
\printead{e3}}
%
%
%
\end{aug}

%
\received{\smonth{7} \syear{2014}}
%
\revised{\smonth{4} \syear{2015}}

%
\begin{abstract}
Bike sharing systems (BSSs) have become a means of sustainable
intermodal transport and are now proposed in many cities worldwide.
Most BSSs also provide open access to their data, particularly to
real-time status reports on their bike stations. The analysis of the
mass of data generated by such systems is of particular interest to BSS
providers to update system structures and policies. This work was
motivated by interest in analyzing and comparing several European BSSs
to identify common operating patterns in BSSs and to propose practical
solutions to avoid potential issues. Our approach relies on the
identification of common patterns between and within systems. To this
end, a model-based clustering method, called FunFEM, for time series
(or more generally functional data) is developed. It is based on a
functional mixture model that allows the clustering of the data in a
discriminative functional subspace. This model presents the advantage
in this context to be parsimonious and to allow the visualization of
the clustered systems. Numerical experiments confirm the good behavior
of FunFEM, particularly compared to state-of-the-art methods. The
application of FunFEM to BSS data from JCDecaux and the Transport for
London Initiative allows us to identify 10 general patterns, including
pathological ones, and to propose practical improvement strategies
based on the system comparison. The visualization of the clustered data
within the discriminative subspace turns out to be particularly
informative regarding the system efficiency. The proposed methodology
is implemented in a package for the R software, named \texttt{funFEM},
which is available on the CRAN. The package also provides a subset of
the data analyzed in this work.
\end{abstract}

%
\end{frontmatter}

\section{Introduction}\label{sec1}
This work was motivated by the will to analyze and compare bike sharing
systems (BSSs) to identify their common strengths and weaknesses. This
type of study is possible because most BSS operators, in dozens of
cities worldwide, provide open access to real-time status reports on
their bike stations (e.g., the number of available bikes, the number of
free bike stands). The implementation of bike sharing systems is one of
the urban mobility services proposed in cities across the world as an
additional means of sustainable intermodal transport. Several studies
[\citet{Froehlich2009,Borgnat11,Vogel2011a,lathia12}] have shown the
usefulness of analyzing\vadjust{\goodbreak} the data collected by BSS operators and city
authorities. A~statistical analysis of these data helps in the
development of new and innovative approaches for a better understanding
of both urban mobility and BSS use. The design of BSSs, the adjustment
of pricing policies and the improvement of system services (e.g.,
redistribution of bikes over stations) can all benefit from this
type of analysis [\citet{Dellolio2011,Lin2011}].

However, the amount of data collected on such systems is often very
large. It is therefore difficult to acquire knowledge using it without
the help of automatic algorithms that extract mobility patterns and
give a synthetic view of the information. This task is usually achieved
in the literature using clustering approaches. In almost all clustering
studies conducted until now, bicycle sharing stations are grouped
according to their usage profiles, thus highlighting the relationships
between time of day, location and usage. In this way, the global
behavior of each station can be efficiently summarized using a few
clusters. These data can be used afterward to analyze the effect of
changing pricing policies or opening new sets of stations [\citet
{lathia12}]. Clustering results can also be used to study the cause of
network imbalance [\citet{Vogel2011a,Vogel2011b,come2014}] and serve
as a first step toward providing automatic reallocation strategies. In
the same way, the clustered results can be used to compare the level of
services reached by the systems of several cities through the
inspection of the proportions of stations that belong to each cluster
in the different cities.

From a methodological point of view, the first attempt in this line of
work was made by \citet{Froehlichmeasuring}, who analyzed a data set
from the Barcelona Bicing system. The data correspond to station
occupancy statistics in the form of free slots, available bikes over
several time frames and other station activity statistics derived from
station occupancy data collected every 5 minutes. The clustering is
performed using a Gaussian mixture model based on features such as the
average number of available bikes at different periods of the day. It
should be noted that such techniques do not really take advantage of
the temporal dynamic of data. In \citet{Froehlich2009}, two types of
clustering are compared, both of which are performed by hierarchical
aggregation. The first one uses activity statistics derived from the
evolution of station occupancy, whereas the second directly uses the
number of available bicycles throughout the day. Other studies, such as
\citet{lathia12}, use similar clustering techniques and data. As in
\citet{Froehlich2009}, each station is described by a time series
vector that corresponds to the normalized available bicycle value of
the station throughout the day. Each element of the feature vector is
therefore equal to the number of available bicycles divided by the
station size. These time series are then smoothed using a moving
average and clustered using a hierarchical agglomerative algorithm [see
page 552 of \citet{Duda01}], with a cosine distance. Another work that
uses the same type of data was proposed by \citet
{Vogel2011a,Vogel2011b}; it uses feature vectors to describe the
stations that come from normalizing arrival and departure counts per\vadjust{\goodbreak}
hour and also handles weekdays and weekends separately. Classical
clustering algorithms, that is, $k$-means, Gaussian mixture
models and sequential information bottleneck (sIB), are then compared.
Finally, \citet{come2014} recently proposed an original approach
considering a generative model based on Poisson mixtures to cluster
stations with respect to hourly usage profiles build from trip data.
The results obtained for the V\'elib' system (Paris) were then analyzed
with respect to the city geography and sociology.

However, all of these works share two limitative characteristics: They
are limited to one BSS (one city), and they do not explicitly model the
functional nature of the data. Indeed, the observed time series are
clustered in those works using either geometric methods based on
distances between time series or by creating features that summarize
the activity in the given periods of the day (and thus omitting the
temporal dynamics of the data). In this work, we aim to go beyond the
analyses made in those works by comparing several European BSSs using a
clustering approach designed for time series data. To this end, we
introduce a novel model-based clustering method devoted to time series
(and, more generally, functional data) that is able to take into
account the nature of the BSS data. The proposed methodology, called
FunFEM, is based on the discriminative functional mixture (DFM) model,
which models the data into a single discriminative functional subspace.
This subspace subsequently allows an insightful visualization of the
clustered data and eases the comparison of systems regarding the
identified patterns. A family of 12 models is also proposed by relaxing
or constraining the main DFM model, allowing it to handle a wide range
of situations. The FunFEM algorithm is proposed for the inference of
the DFM models, and model selection can be performed either by BIC or
the ``slope heuristic.'' In addition, the selection of the most
discriminative basis functions can be made afterward by introducing
sparsity through a $\ell_1$-type penalization. The comparison of 8
European BSS using FunFEM allows us to identify pathological and
healthy patterns in the system dynamic and to propose practical
improvement strategies based on the most efficient systems.

The paper is organized as follows. Section~\ref{sec2} presents the BSS data used
to analyze and compare several European bike sharing systems. Section~\ref{sec3}
introduces the DFM model, its model family and the FunFEM algorithm.
The model choice and selection of the discriminative functions are also
discussed in Section~\ref{sec3}. Numerical experiments on simulated and
benchmark data sets are then presented in Section~\ref{sec4} to validate the
proposed approach. Section~\ref{sec5} presents the analyses and comparisons of 8
bike sharing systems using the FunFEM algorithm. Based on the
comparison results, recommendations to BSS providers and city planners
are made. Finally, Section~\ref{sec6} provides concluding remarks.

\begin{table}
\tablewidth=250pt
\caption{Summary statistics for the eight bike
sharing systems involved in the study}
\label{tab:bsslist}
\begin{tabular*}{250pt}{@{\extracolsep{\fill}}lcc@{}}
\hline
\textbf{City} & \textbf{Stations} & \textbf{Bikes}\\
\hline
Paris & 1230 & 18,000\\
London & \phantom{0}740 & \phantom{0.}9500\\
Lyon & \phantom{0}345 & \phantom{0.}3200 \\
Bruxelles & \phantom{0}330 & \phantom{0.}3800\\
Valencia & \phantom{0}280 & \phantom{0.}2400 \\
Sevilla & \phantom{0}260 & \phantom{0.}2150 \\
Marseille & \phantom{0}120 & \phantom{00.}650 \\
Nantes & \phantom{0}102 & \phantom{00.}880 \\
\hline
\end{tabular*}
\end{table}
%

\begin{figure}

\includegraphics{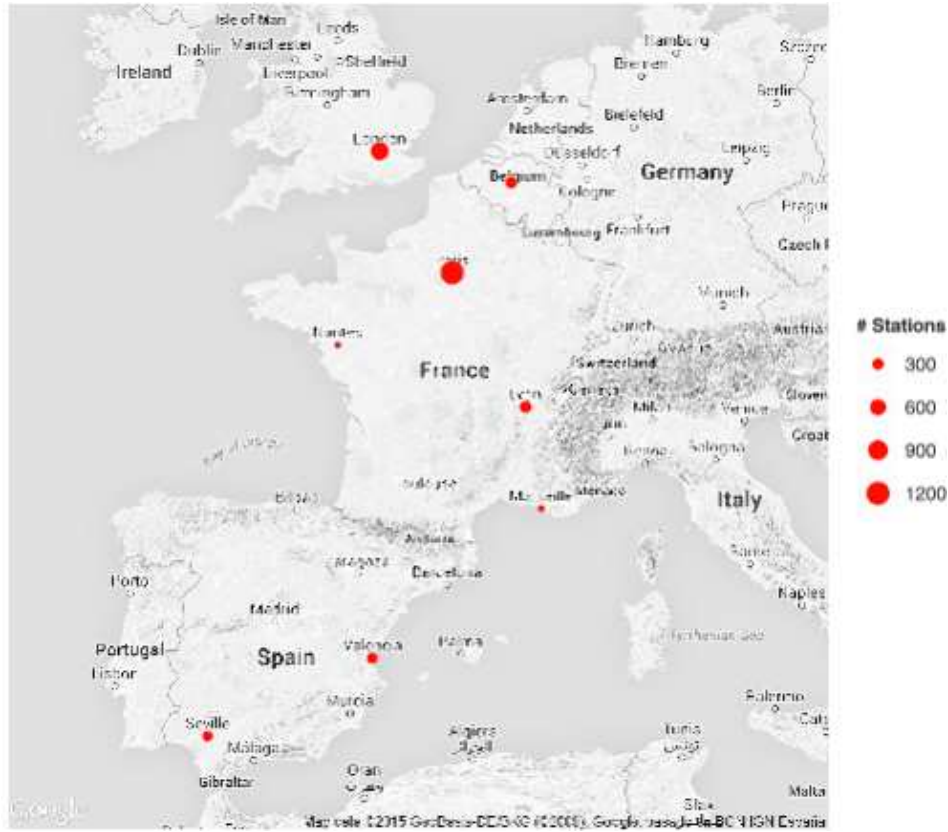}

\caption{Map of the eight European bike sharing
systems involved in the study. The dot size denotes the system size.}\label{fig:Map}
\end{figure}

\section{The BSS data}\label{sec2}

In this work we want to analyze station occupancy data collected over
the course of one month on eight bike sharing systems in Europe. The
data were collected over 5 weeks, between February, 24 and March, 30,
2014. Table~\ref{tab:bsslist} lists the BSSs included in this study
and some summary statistics on the systems. Figure~\ref{fig:Map}
visualizes the locations of the studied systems. The cities were chosen
to cover different cases in terms of the geographic positions of the
city (south/north of Europe) and to cover a range of system sizes,
from small-scale systems, such as Nantes, to much larger systems, such
as Paris.

The station status information, in terms of available bikes and docks,
were downloaded every hour during the study period for the seven
systems from the open-data APIs provided by the JCDecaux
company\footnote{The real-time data are available at
\url{https://developer.jcdecaux.com/} (with an api key).} and by the
Transport for London initiative.\footnote{The real-time data are
available at \url{https://www.tfl.gov.uk/info-for/open-data-users/} (with an
api key).} To accommodate the varying stations sizes (in terms of the
number of docking points), we normalized the number of available bikes
by the station size and obtained a loading profile for each station.
The final data set contains 3230 loading profiles, one per station,
sampled at 1448 time points. Notice that the sampling is not perfectly
regular; there is an hour, on average, between the two sample points.

\begin{figure}

\includegraphics{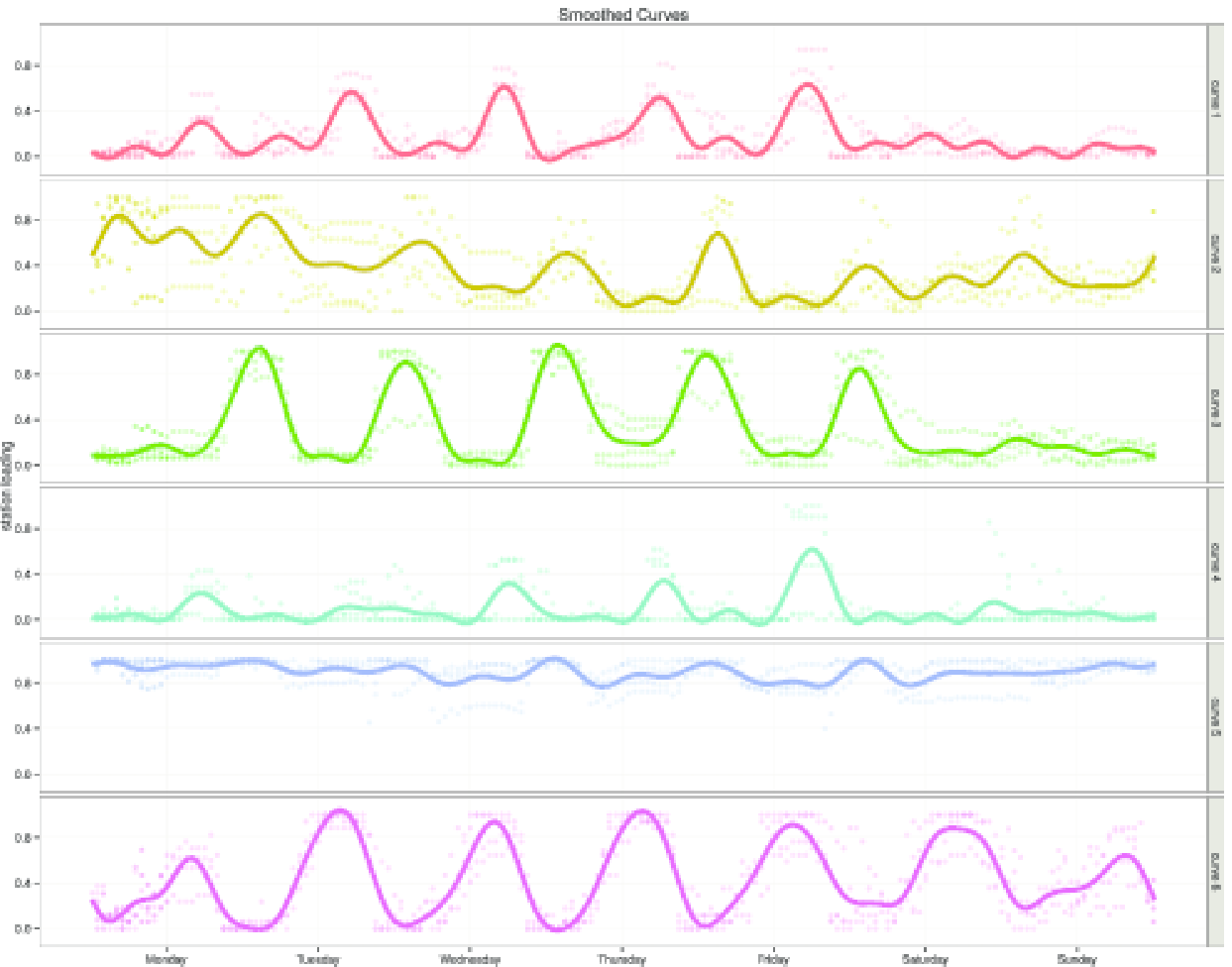}

\caption{Some examples of smoothed station
profiles, with the corresponding observations. One month of
observations is depicted here using a period of one week.} \label{fig:smoothing}
\end{figure}

The daily and weekly habits of inhabitants introduce a periodic
behavior in the BSS station loading profiles, with a natural period of
one week. It is then natural to use a Fourier basis to smooth the
curves, with basis functions corresponding to sine and cosine functions
of periods equal to fractions of this natural period of the data. Using
such a procedure, the profiles of the 3230 stations were projected on a
basis of 41 Fourier functions (see Section~\ref{sec3} for details); the smoothed
curves obtained for 6 different stations are depicted in Figure~\ref{fig:smoothing},
together with the curve samples. A typical periodic
behavior is clearly visible in this figure for some stations. Some
other stations exhibit, however, a less clear pattern, such as curves
2, 4 and 5. Our study aims, therefore, to identify the different
patterns hidden in the data using functional clustering and to use them
to compare the eight studied systems.

\section{The discriminative functional mixture model}\label{sec3}

From a theoretical point of view, the aim of this work is to cluster a
set of observed curves $\{x_{1},\ldots,x_{n}\}$ (the loading function of
the bike stations) into $K$ homogenous groups (or clusters), allowing
for the analysis of the studied process. After a short review of
related works in functional data clustering, this section introduces a
latent functional model that adapts the model of \citet
{Bouveyron12FEM} proposed in the multivariate case to functional data.
An original inference algorithm for the functional model is then
proposed, subsequently allowing for the clustering of the curves. The
model choice and variable selection are also discussed.

\subsection{Related work in functional clustering}\label{sec3.1}

This work is rooted in the recent advances in functional data analysis
that have contributed to the development of efficient clustering
techniques specific to functional data. One of the earlier works in
that domain was by \citet{Jam2003}, who defined an approach that is
particularly effective for sparsely sampled functional data. This
method, called fclust, considers that the basis expansion coefficients
of curves into a spline basis are distributed according to a mixture of
Gaussians with cluster-specific means and common variances. The use of
a spline basis is convenient when the curves are regular but are not
appropriate for peak-like data, for instance, the data encountered in
mass spectrometry. For this reason, \citet{Gia2012} recently proposed a
Gaussian model on a wavelet decomposition of curves. This approach
allows for addressing a wider range of functional shapes than splines.
An interesting approach has also been considered by \citet{Sam2011},
who assume that curves arise from a mixture of regressions based on
polynomial functions, with possible regime changes at each instant of
observation. Let us also mention the work of \citet{Fru2008}, who have
built a specific clustering algorithm based on parametric time series
models. \citet{Bou2011} extended the high-dimensional data clustering
(HDDC) algorithm [\citet{Bouveyron07b}] to the functional case. The
resulting model assumes a parsimonious cluster-specific Gaussian
distribution for the basis expansion coefficients. More recently, \citet
{Jac2013} proposed a model-based clustering built on the approximation
of the notion of density for functional variables, extended to
multivariate functional data in \citet{Jac2014}. These models assume
that the functional principal component scores of curves have a
Gaussian distribution whose parameters are cluster-specific. Bayesian
approaches have also been proposed: For example, \citet{Hea2006}
consider that the basis expansion coefficients are distributed as a
mixture of Gaussians whose variances are modeled by an Inverse-Gamma
distribution. Further, \citet{Ray2006} propose a nonparametric Bayes
wavelet model for curve clustering based on a mixture of Dirichlet processes.

\subsection{Transformation of the observed curves}\label{sec3.2}
Let us first assume that the observed curves $\{x_{1},\ldots,x_{n}\}$ are
independent realizations of a $L_{2}$-continuous stochastic process
$X=\{X(t)\}_{t\in[0,T]}$ for which the sample paths, that
is, the observed curves, belong to $L_{2}[0,T]$. In practice, the
functional expressions of the observed curves are not known, and we
have access only to the discrete observations $x_{ij}=x_{i}(t_{is})$ at
a finite set of ordered times $\{t_{is}:s=1,\ldots,m_{i}\}$. It is
therefore necessary to first reconstruct the functional form of the
data from their discrete observations. A common way to do this is to
assume that the curves belong to a finite dimensional space spanned by
a basis of functions [see, e.g., \citet{Ram2005}]. Let us
therefore consider such a basis $\{\psi_{1},\ldots,\psi_{p}\}$ and
assume that the stochastic process $X$ admits the following basis expansion:
%
\begin{equation}
X(t)=\sum_{j=1}^{p}\gamma_{j}(X)
\psi_{j}(t),\label{eq:X}
\end{equation}
where $\gamma=(\gamma_{1}(X),\ldots,\gamma_{p}(X))$ is a random vector
in $\mathbb{R}^{p}$, and the number $p$ of basis functions is assumed
to be fixed and known. The basis expansion of each observed curve
$x_{i}(t)=\sum_{j=1}^{p}\gamma_{ij}\psi_{j}(t)$ can be estimated by
an interpolation procedure [see \citet{Esc2005}, e.g.] if the
curves are observed without noise or by least squares smoothing if they
are observed with error:
\[
x_{i}^{\mathrm{obs}}(t_{is})=x_{i}(t_{is})+
\varepsilon_{is},\qquad s=1,\ldots,m_{i}.
\]
The latter option is used in the present work. In this case, the basis
coefficients of each sample path $x_{i}$ are approximated by
\[
\widehat{\gamma}_{i}= \bigl(\Theta_{i}^{\prime}
\Theta_{i} \bigr)^{-1}\Theta_{i}^{\prime}X_{i}^{\mathrm{obs}},
\]
with $\Theta_{i}=(\psi_{j}(t_{is}))_{1\leq j\leq n,1\leq s\leq
m_{i}}$ and $X_{i}^{\mathrm{obs}}=(x_{i}^{\mathrm{obs}}(t_{i_{1}}),\ldots,x_{i}^{\mathrm{obs}}(t_{i_{m_{i}}}))^{\prime}$.

\subsection{The model}\label{sec3.3}

The goal is to cluster the observed curves $\{x_{1},\ldots,x_{n}\}$
into $K$ homogeneous groups. Let us assume that there exists an unobserved
random variable $Z=(Z_{1},\ldots,Z_{K})\in\{0,1\}^{K}$ indicating
the group membership of $X$: $Z_{k}$ is equal to 1 if $X$ belongs
to the $k$th group and 0 otherwise. The clustering task aims therefore
to predict the value $z_{i}=(z_{i1},\ldots,z_{iK})$ of $Z$ for each
observed curve $x_{i}$.

Let $F[0,T]$ be a latent subspace of $L_{2}[0,T]$ assumed to be
the most discriminative subspace for the $K$ groups spanned by a
basis of $d$ basis functions $\{\varphi_{j}\}_{j=1,\ldots,d}$ in $L_{2}[0,T]$,
with $d<K$ and $d<p$. The assumption $d<K$ is motivated by the fact
that a subspace of $d=K-1$ dimensions is sufficient to discriminate $K$
groups [\citet{Fisher36,Fukunaga90}]. The basis $\{\varphi_{j}\}
_{j=1,\ldots,d}$ is obtained
from $\{\psi_{j}\}_{j=1,\ldots,p}$ through a linear transformation
$\varphi_{j}=\sum_{\ell=1}^{p}u_{j\ell}\psi_{\ell}$
such that the $p\times d$ matrix $U=(u_{j\ell})$ is orthogonal.
Let $\{\lambda_{1},\ldots,\lambda_{n}\}$ be the latent expansion coefficients
of the curves $\{x_{1},\ldots,x_{n}\}$ on the basis $\{\varphi_{j}\}_{j=1,\ldots,d}$.
These coefficients are assumed to be independent realizations of a
latent random vector $\Lambda\in\mathbb{R}^{d}$. The relationship
between the bases $\{\varphi_{j}\}_{j=1,\ldots,d}$ and $\{\psi_{j}\}_{j=1,\ldots,p}$
suggests that the random vectors $\Gamma$ and $\Lambda$ are linked
through the following linear transformation:
%
\begin{equation}
\Gamma=U\Lambda+\varepsilon,
\end{equation}
where $\varepsilon\in\mathbb{R}^{p}$ is an independent and random
noise term.

Let us now make distributional assumptions on the random vectors
$\Lambda$ and $\varepsilon$. First, conditionally on $Z$, $\Lambda$
is assumed to be
distributed according to a multivariate Gaussian density:
%
\begin{equation}
\Lambda_{|Z=k}\sim\mathcal{N}(\mu_{k},\Sigma_{k}),
\end{equation}
where $\mu_{k}$ and $\Sigma_{k}$ are, respectively, the mean and the
covariance
matrix of the $k$th group. Second, $\varepsilon$ is also assumed to be
distributed according to a multivariate Gaussian density:
%
\begin{equation}
\varepsilon\sim\mathcal{N}(0,\Xi).
\end{equation}
With these distributional assumptions, the marginal distribution of
$\Gamma$ is a mixture of Gaussians:
%
\begin{equation}
p(\gamma)=\sum_{k=1}^{K}\pi_{k}
\phi\bigl(\gamma;U\mu_{k},U^{t}\Sigma _{k}U+\Xi
\bigr),
\end{equation}
where $\phi$ is the standard Gaussian density function, and $\pi
_{k}=P(Z=k)$ is the prior probability of the $k$th group.

\begin{figure}

\includegraphics{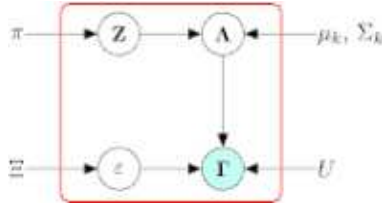}

\caption{Graphical representation for the model
$\mathrm{DFM}_{[\Sigma_{k}\beta]}$.} \label{fig:GraphModel}
\end{figure}

We finally assume that the noise covariance matrix $\Xi$ is such that
$\Delta_{k}=\operatorname{cov}(W^{t}\Gamma|Z=k)=W^{t}\Sigma_{k}W$ has the
following form:
%
\begin{eqnarray}
\small{\Delta_{k}=\left(\begin{array}{c@{}c}
\begin{array}{|ccc|}
\hline  &  & \\
 & \Sigma_{k} & \\
 &  & \\
\hline \end{array} & \mathbf{0}\\
\mathbf{0} & \begin{array}{|ccc|}
\hline \beta &  & 0\\
 & \ddots & \\
0 &  & \beta
\\\hline \end{array}
\end{array}\right)\begin{array}{cc}
\left.\begin{array}{c}
\\
\\
\end{array}\right\}  & d\vspace*{10pt}\\
\left.\begin{array}{c}
\\
\\
\\
\end{array}\right\}  & p-d
\end{array}\label{eqdelta}}\label{eq:Delta}
\end{eqnarray}
with $W=[U,V]$, where $V$ is the orthogonal complement of $U$. With
this notation, and from a practical point of view, one can say that the
variance of the actual data of the $k$th group is therefore modeled by
$\Sigma_{k}$, whereas the parameter $\beta$ models the variance
of the noise outside the functional subspace. This model is referred to
in the sequel as $\mathrm{DFM}_{[\Sigma_{k}\beta]}$, and
Figure~\ref{fig:GraphModel} summarizes the modeling.

%
\begin{table}
\caption{Number of free parameters in covariance matrices
when $d=K-1$ for the DFM models}
\label{Tab:models}
\begin{tabular*}{\textwidth}{@{\extracolsep{\fill}}lccc@{}}
\hline
\multicolumn{1}{@{}l}{\textbf{Model}} & \multicolumn{1}{c}{$\bolds{\Sigma_{k}}$} & \multicolumn{1}{c}{$\bolds{\beta_{k}}$} & \multicolumn{1}{c@{}}{\textbf{Nb. of variance parameters}}
\\
\hline
$\mathrm{DFM}_{[\Sigma_{k}\beta_{k}]}$ & Free & Free &
$(K-1)(p-K/2)+K^{2}(K-1)/2+K$ \\
$\mathrm{DFM}_{[\Sigma_{k}\beta]}$ & Free & Common &
$(K-1)(p-K/2)+K^{2}(K-1)/2+1$\\
$\mathrm{DFM}_{[\Sigma\beta_{k}]}$ & Common & Free &
$(K-1)(p-K/2)+K(K-1)/2+K$ \\
$\mathrm{DFM}_{[\Sigma\beta]}$ & Common & Common &
$(K-1)(p-K/2)+K(K-1)/2+1$\\
$\mathrm{DFM}_{[\alpha_{kj}\beta_{k}]}$ & Diagonal & Free &
$(K-1)(p-K/2)+K^{2}$ \\
$\mathrm{DFM}_{[\alpha_{kj}\beta]}$ & Diagonal & Common &
$(K-1)(p-K/2)+K(K-1)+1$ \\
$\mathrm{DFM}_{[\alpha_{k}\beta_{k}]}$ & Spherical & Free &
$(K-1)(K-1)(p-K/2)+2K$ \\
$\mathrm{DFM}_{[\alpha_{k}\beta]}$ & Spherical & Common &
$(K-1)(p-K/2)+K+1$ \\
$\mathrm{DFM}_{[\alpha_{j}\beta_{k}]}$ & Diagonal \& Common & Free &
$(K-1)(p-K/2)+(K-1)+K$ \\
$\mathrm{DFM}_{[\alpha_{j}\beta]}$ & Diagonal \& Common & Common &
$(K-1)(p-K/2)+(K-1)+1$ \\
$\mathrm{DFM}_{[\alpha\beta_{k}]}$ & Spherical \& Common & Free &
$(K-1)(p-K/2)+K+1$ \\
$\mathrm{DFM}_{[\alpha\beta]}$ & Spherical \& Common & Common &
$(K-1)(p-K/2)+2$ \\
\hline
\end{tabular*}
\end{table}

\subsection{A family of discriminative functional model}\label{sec3.4}

Starting with the model $\mathrm{DFM}_{[\Sigma_{k}\beta]}$ and
following the strategy of \citet{Fraley99}, several submodels can be
generated by applying constraints on the parameters of the matrix
$\Delta_{k}$. For instance, it is first possible to relax the
constraint that the noise variance is common across groups. This
generates the model $\mathrm{DFM}_{[\Sigma_{k}\beta_{k}]}$, which is
the more general model of the family. It is also possible to constrain
this new model such that the covariance matrices $\Sigma_{1},\ldots
,\Sigma_{K}$ in the latent space are common across groups. This
submodel will be referred to as $\mathrm{DFM}_{[\Sigma\beta_{k}]}$.
Similarly, in each group, $\Sigma_{k}$ can be assumed to be diagonal,
that is, $\Sigma_{k}=\operatorname{diag}(\alpha_{k1},\ldots
,\alpha_{kd})$, and this submodel will be referred to as $\mathrm
{DFM}_{[\alpha_{kj}\beta_{k}]}$. The variance within the latent
subspace $F$ can also be assumed to be isotropic for each group, and
the associated submodel is $\mathrm{DFM}_{[\alpha_{k}\beta_{k}]}$.
Following this strategy, $12$ different DFM models can be enumerated,
and an overview of them is proposed in Table~\ref{Tab:models}. The
table also provides, for each model, the number of variance parameters
to estimate as a function of the number $K$ of groups and the number
$p$ of basis functions.
One can note that the models turn out to be particularly parsimonious
because their complexity is a linear function of $p$, whereas most
model-based approaches usually have a complexity that is a quadratic
function of $p$.

%
\subsection{Model inference: The FunFEM algorithm}\label{sec3.5}

Because the group memberships $\{z_{1},\ldots,z_{n}\}$ of the curves are
unknown, the direct maximization of the likelihood associated with the
model described above is intractable. In such a case, a classical
solution for model inference is to use the EM algorithm. Here, however,
the use of the EM algorithm is prohibited due to the particular nature
of the functional subspace $F$. Indeed, maximizing the likelihood over
the subspace orientation matrix $U$ is equivalent to maximizing the
projected variance, and it yields the functional principal component
analysis (fPCA) subspace. Because $F$ is here assumed to be the most
discriminative subspace, $U$ has to be estimated separately, and we
therefore propose the algorithm described hereafter and named FunFEM.
The FunFEM algorithm alternates, at iteration $q$, over the three
following steps:

\textit{The F step}.
Let us first suppose that at iteration $q$, the posterior probabilities
$t_{ik}^{(q)}=E[z_{ik}|\gamma_{i},\theta^{(q-1)}]$ are known (they
have been estimated in the E step of iteration $q-1$). The F step aims
therefore to determine, conditionally on the $t_{ik}^{(q)}$, the
orientation matrix $U$ of the discriminative latent subspace $F$ in
which the $K$ clusters are best separated. Following the original idea
of Fisher (\citeyear{Fisher36}), the functional subspace $F$ should be such that the
variance within the groups should be minimal, whereas the variance
between groups should be maximal.
Let $\mathbf{C}$ be the covariance operator of $X$ with kernel
\[
C(t,s)=\mathbb{E} \bigl[\bigl(X(t)-m(t)\bigr) \bigl(X(s)-m(s)\bigr) \bigr],
\]
and $\mathbf{B}$ be the integral between-cluster covariance operator
with kernel
\[
B(t,s)=\mathbb{E} \bigl[\mathbb{E}\bigl[X(t)-m(t)|Z\bigr]\mathbb {E}
\bigl[X(s)-m(s)|Z\bigr] \bigr],
\]
where $m(t)=\mathbb{E}[X(t)]$. In the following, and without a loss of
generality, the curves are assumed to be centered, that is,
$m(t)=0$. The operator $\mathbf{B}$ can thus be rewritten as follows:
\begin{eqnarray*}
B(t,s) & = & \mathbb{E} \bigl[\mathbb{E}\bigl[X(t)|Z\bigr]\mathbb {E}
\bigl[X(s)|Z\bigr] \bigr]
\\
& = & \mathbb{E} \Biggl[\sum_{k=1}^{K}
\mathbf{1}_{\{Z=k\}}\mathbb {E}\bigl[X(t)|Z=k\bigr]\sum
_{\ell=1}^{K}\mathbf{1}_{\{Z=\ell\}}\mathbb {E}
\bigl[X(s)|Z=\ell\bigr] \Biggr]
\\
& = & \sum_{k=1}^{K}P(Z=k)\mathbb{E}
\bigl[X(t)|Z=k\bigr]\mathbb{E}\bigl[X(s)|Z=k\bigr].
\end{eqnarray*}
The Fisher criterion, in the functional case and the supervised setting
[\citet{Pre2007}], looks for the discriminative function $u\in L_{2}[0,T]$
which is solution of
%
\begin{equation}
\max_{u}\frac{\operatorname{Var}(\mathbb{E}[\Phi(X)|Z])}
{\operatorname{Var}(\Phi(X))},\label
{eq:FisherCrit}
\end{equation}
where $\Phi(X)=\int_{[0,T]}X(t)u(t)\,dt$ is the projection of $X$ on
the discriminative function $u$. Let us recall that we consider here
the unsupervised setting, and $Z$ is an unobserved variable. The
solution of (\ref{eq:FisherCrit}) is the eigenfunction $u$ associated
with the largest eigenvalue $\eta\in\mathbb{R}$ of the following
generalized eigenproblem:
%
\begin{eqnarray}\label
{eq:GEProblem}
\mathbf{B}u & = & \eta\mathbf{C}u,
\nonumber
\\[-8pt]
\\[-8pt]
\nonumber
\int_{[0,T]}B(t,s)u(s)\,ds & = & \eta\int_{[0,T]}C(t,s)u(s)\,ds,
\end{eqnarray}
under the constraint $<u,\mathbf{C}u>_{L_{2}[0,T]}=1$.
The estimator for $C(t, s)$ from the sample $\{x_{1},\ldots,x_{n}\}$,
expanded on the basis $ (\psi_{j} )_{j=1,\ldots,p}$, is
\begin{eqnarray*}
\hat{C}(t,s) & = & \frac{1}{n}\sum_{i=1}^{n}
\Biggl(\sum_{j=1}^{p}\gamma _{ij}
\psi_{j}(t)\Biggr) \Biggl(\sum_{j=1}^{p}
\gamma_{ij}\psi_{j}(s)\Biggr)
\\
& = & \frac{1}{n}\Psi'(t)\bolds{\Gamma}'\bolds{
\Gamma}\Psi(s),
\end{eqnarray*}
where $\bolds{\Gamma}=(\gamma_{ij})_{i, j}$ is the $n\times p$-matrix
of basis expansion coefficients and $\Psi(s)$ is the $p$-vector of the
basis functions $\psi_{j}(s)$ ($1\leq i\leq n$ and $1\leq j\leq p$).
Because the variable $Z$ is unobserved, $B(t, s)$ has to be estimated
conditionally on the posterior probabilities
$t_{ik}^{(q-1)}=E[z_{ik}|\gamma_{i},\theta^{(q-1)}]$ obtained from
the E step at iteration $q-1$:
\begin{eqnarray*}
\hat{B}^{(q)}(t,s) & = & \sum_{k=1}^{K}
\frac{n_{k}^{(q-1)}}{n} \Biggl(\frac{1}{n_{k}^{(q-1)}}\sum_{i=1}^{n}t_{ik}^{(q-1)}x_{i}(t)
\Biggr) \Biggl(\frac{1}{n_{k}^{(q-1)}}\sum_{i=1}^{n}t_{ik}^{(q-1)}x_{i}(s)
\Biggr)
\\
& = & \frac{1}{n}\sum_{k=1}^{K}
\frac{1}{n_{k}^{(q-1)}} \Biggl(\sum_{i=1}^{n}t_{ik}^{(q-1)}
\sum_{j=1}^{p}\gamma_{ij}
\psi_{j}(t) \Biggr) \Biggl(\sum_{i=1}^{n}t_{ik}^{(q-1)}
\sum_{j=1}^{p}\gamma_{ij}\psi
_{j}(s) \Biggr),
\end{eqnarray*}
and in a matrix form:
\[
\hat{B}^{(q)}(t,s)=\frac{1}{n}\Psi'(t)\bolds{
\Gamma}'\mathbf{T}\mathbf {T}'\bolds{\Gamma}\Psi(s),
\]
with $n_{k}^{(q-1)}=\sum_{i=1}^{n}t_{ik}^{(q-1)}$ and $\mathbf{T}=
(\frac{t_{ik}^{(q-1)}}{\sqrt{n_{k}^{(q-1)}}} )_{i, k}$ is a
$n\times K$-matrix.
Assuming that the discriminative function $u$ can be decomposed in the
same basis as the observed curves,
%
\begin{equation}
u(t)=\sum_{j=1}^{p}\nu_{j}
\psi_{j}(t)=\Psi'(t)\bolds{\nu},\label{eq:X-1}
\end{equation}
the generalized eigenproblem (\ref{eq:GEProblem}) becomes
\[
\int_{[0,T]}\frac{1}{n}\Psi'(t)\bolds{
\Gamma}'\mathbf{T}\mathbf {T}'\bolds{\Gamma}\Psi(s)
\Psi'(s)\bolds{\nu}\,ds=\eta\int_{[0,T]}
\frac{1}{n}\Psi'(t)\bolds{\Gamma}'\bolds{\Gamma}
\Psi (s)\Psi'(s)\bolds{\nu}\,ds,
\]
which is equivalent to
\[
\frac{1}{n}\Psi'(t)\bolds{\Gamma}'\mathbf{T}
\mathbf{T}'\bolds{\Gamma }W\bolds{\nu}=\eta\frac{1}{n}
\Psi'(t)\bolds{\Gamma}'\bolds{\Gamma }W\bolds{\nu},
\]
with $\mathbf{W}=\int_{[0,T]}\Psi(s)\Psi'(s)\,ds$. Because this equality
holds for all $t\in[0,T]$, we have
\[
\bolds{\Gamma}'\mathbf{T}\mathbf{T}'\bolds{\Gamma}W
\bolds{\nu}=\eta \bolds{\Gamma}'\bolds{\Gamma}W\bolds{\nu},
\]
or, equivalently,
%
\begin{equation}
\bigl(\bolds{\Gamma}'\bolds{\Gamma}W\bigr)^{-1}\bolds{
\Gamma}'\mathbf{T}\mathbf {T}'\bolds{\Gamma}W\bolds{
\nu}=\eta\bolds{\nu}.\label{eq:GEP2}
\end{equation}
Finally, the basis expansion coefficient $\nu=(\nu_{1},\ldots,\nu
_{p})'$ of the discriminative function $u$ is the eigenvector of the
above generalized eigenproblem associated with the largest eigenvalue.
Once the first discriminative function, let us say $u_{1}$, is
determined, the second discriminative function is obtained by solving
the generalized eigenproblem (\ref{eq:GEP2}) in the complementary
space of $u_{1}$. This procedure is recursively applied until the $d$
discriminative functions $\{u_{1},\ldots,u_{d}\}$ are obtained. The basis
expansion coefficients $\nu_{j}^{(q)}=(\nu_{j1}^{(q)},\ldots,\nu
_{jp}^{(q)})'$, $j=1,\ldots, d$ of the estimated discriminative
functions are gathered in the $p\times d$ matrix $U^{(q)}= (\nu
_{j\ell}^{(q)} )_{j,\ell}$.

\textit{The M step}.
Following the classical scheme of the EM algorithm, this step aims to
maximize, conditionally on the orientation matrix $U^{(q)}$ obtained
from the previous step, the conditional expectation of the complete
data log-likelihood $Q(\theta;\theta^{(q-1)})=E [\ell(\theta
;\bolds{\Gamma},z_{1},\ldots,z_{n})|\bolds{\Gamma},\theta^{(q-1)} ]$:
\begin{eqnarray*}
&&Q\bigl(\theta;\theta^{(q-1)}\bigr) \\
&&\qquad =  -\frac{1}{2}\sum
_{k=1}^{K}n_{k}^{(q-1)} \Biggl[\log|
\Sigma_{k}|+(p-d)\log(\beta )-2\log(\pi_{k})+p\log(2\pi)
\\
& &\qquad\quad{} +\frac{1}{n_{k}^{(q-1)}}\sum_{i=1}^{n}t_{ik}^{(q-1)}(
\gamma_{i}-\mu_{k})^{t}U^{(q)}\Delta
_{k}^{-1}U^{(q)}{}^{t}(
\gamma_{i}-\mu_{k}) \Biggr]
\\
&&\qquad =  -\frac{1}{2}\sum_{k=1}^{K}n_{k}^{(q-1)}
\Biggl[\log|\Sigma _{k}|+(p-d)\log(\beta)-2\log(\pi_{k})+p
\log(2\pi)
\\
& &\qquad\quad{} +\operatorname{trace}\bigl(\Sigma _{k}^{-1}U^{(q)}{}^{t}C_{k}U^{(q)}
\bigr)+\frac{1}{\beta} \Biggl(\operatorname {trace}(C_{k})-\sum
_{j=1}^{d}\nu_{j}^{(q)t}C_{k}
\nu_{j}^{(q)} \Biggr) \Biggr],
\end{eqnarray*}
where $\theta=(\pi_{k},\mu_{k},\Sigma_{k},\beta)_{k}$, for $1\leq
k\leq K$, and $C_{k}=\frac{1}{n_{k}^{(q-1)}}\sum_{i=1}^{n}t_{ik}^{(q-1)}(\gamma_{i}-\mu_{k}^{(q-1)})(\gamma_{i}-\mu
_{k}^{(q-1)})^{t}$.
The maximization of $Q(\theta;\theta^{(q-1)})$, according to $\pi
_{k},\mu_{k},\Sigma_{k}$ and $\beta$, yields the following updates
for model parameters:
\begin{itemize}
\item$\pi_{k}^{(q)}=n_{k}^{(q-1)}/n$,
\item$\mu_{k}^{(q)}=\frac{1}{n_{k}^{(q-1)}}\sum_{i=1}^{n}t_{ik}^{(q-1)}U^{(q)t}\gamma_{i}$,
\item$\Sigma_{k}^{(q)}=U^{(q)t}C_{k}^{(q)}U^{(q)}$,
\item$\beta^{(q)}= (\operatorname{trace}(C^{(q)})-\sum_{j=1}^{d}u_{j}^{(q)t}C^{(q)}u_{j}^{(q)} )/ (p-d )$.
\end{itemize}
Updated formula for other models of the family can be easily obtained
from \citet{Bouveyron12FEM}.

\textit{The E step}.
This last step reduces to update, at iteration $q$, the posterior
probabilities $t_{ik}^{(q)}=E[z_{ik}|\gamma_{i},\theta^{(q)}]$. Let
us also recall that $t_{ik}^{(q)}$ is also the posterior probability
$P(z_{ik}=1|\gamma_{i},\theta^{(q)})$ that the curve $x_{i}$ belongs
to the $k$th component of the mixture under the current model. Using
Bayes' theorem, the posterior probabilities $t_{ik}^{(q)}$,
$i=1,\ldots,n$, $k=1,\ldots,K$, can be expressed as follows:
%
\begin{equation}
t_{ik}^{(q)}=\frac{\pi_{k}^{(q)}\phi(\gamma_{i},\theta
_{k}^{(q)})}{\sum_{l=1}^{K}\pi_{l}^{(q)}\phi(\gamma_{i}|\theta
_{l}^{(q)})},\label{eq:tik_GMM-1}
\end{equation}
where $\theta_{k}^{(q)}=(\pi_{k}^{(q)},\mu_{k}^{(q)},\Sigma
_{k}^{(q)},\beta^{(q)})$ is the set of parameters for the $k$th
component updated in the M step.

\subsection{Model selection}\label{sec3.6}
We now discuss both the choice of the most appropriate model within the
family and the problem of selecting the number $K$ of groups and the
intrinsic dimension $d$. On the one hand, it is first of interest to
select the model of the DFM family that is the most appropriate to
model the data at hand. On the other hand, the problem of selecting $K$
and $d$ can be, in fact, recast as a model selection problem. The idea
here is to consider, for instance, a DFM model with $K=2$ and the same
DFM model with $K=3$ as two different models among which one wants to
choose. Thus, because a model is defined by its parametrization, its
number of components $K$ and its intrinsic dimensionality $d$, model
selection criteria allow us to select the best combination of those
three features required for modeling the data.

Classical tools for model selection include the AIC [\citet{Akaike74}]
and BIC [\citet{Schwarz78}] criteria, which penalize the log-likelihood
$\ell(\hat\theta)$ as follows, for model $\mathcal{M}$:
%
\begin{equation}
\operatorname{AIC}(\mathcal{M}) = \ell(\hat\theta) - \xi(\mathcal {M}),\qquad \operatorname{BIC}(
\mathcal{M}) = \ell(\hat\theta) - \frac{\xi
(\mathcal{M})}{2}\log(n),
\end{equation}
where $\xi(\mathcal{M})$ is the number of free parameters of the
model, and $n$ is the number of observations.
The value of $\xi(\mathcal{M})$ is, of course, specific to the model
selected by the practitioner (cf. Table~\ref{Tab:models}).
Although penalized likelihood criteria are widely used, AIC and BIC are
also known to be less efficient in practical situations than in
simulated cases. In particular, the required regularity conditions are
not fully satisfied in the mixture framework [\citet
{Lindsay95,Lindsay2008}] and, hence, the criteria might not be appropriate.

To overcome this drawback, \citet{birge2007minimal} recently proposed a
data-driven technique, called the ``slope heuristic,'' to calibrate the
penalty involved in penalized criteria. The slope heuristic was first
proposed in the context of Gaussian homoscedastic least squares
regression and was then used in different situations, including
model-based clustering. \citet{birge2007minimal} showed that there
exists a minimal penalty and that considering a penalty equal to twice
this minimal penalty allows for approximating the oracle model in terms
of risk. The minimal penalty is, in practice, estimated by the slope of
the linear part when plotting the log-likelihood $\ell(\hat\theta)$
with regard to the number of model parameters (or model dimension). The
criterion associated with the slope heuristic is therefore defined by
%
\begin{equation}
\operatorname{SHC}(\mathcal{M}) = \ell(\hat\theta) - 2\hat{s}\xi (\mathcal{M}),
\end{equation}
where $\hat{s}$ is the slope of the linear part of $\ell(\hat\theta)$.
A detailed overview and advice for implementation are provided in
\citet{baudry2012slope}. Section~\ref{sec3} proposes a comparison of the slope
heuristic with classical model selection criteria. In Section~\ref{sec4} the
slope heuristic criterion is used for the model selection for the BSS data.

\subsection{Selection of discriminative basis functions}\label{sec3.7}

Another advantage of the proposed modeling is the possibility of using
the discriminative subspace to select the relevant basis functions for
discriminating between the groups. Indeed, the functional subspace $F$
allows for determining the discriminative basis functions through the
loading matrix $U$, which contains the coefficients of the linear
relation that links the basis functions with the subspace $F$. It is
therefore expected that basis functions associated with large absolute
values of $U$ are particularly relevant for discriminating between the groups.
An intuitive way to identify the discriminative basis functions would
be to keep only large absolute loading variables by, for instance,
thresholding. Although this approach is commonly used in practice, it
has been particularly criticized by \citet{Cadima1995} because it
induces some misleading information. Here, we propose selecting the
discriminative basis functions by constraining the optimization
problem (\ref{eq:FisherCrit}) of the F step such that the loading
matrix $U$ is sparse (i.e., such that $U$ contains as many
zeros as possible). To this end, we follow the approach proposed
by \citet{Bouveyron14a}, who rewrite the constrained Fisher criterion
as a $\ell_{1}$-penalized regression problem. We therefore use their
algorithm [Algorithm 2 of \citet{Bouveyron14a}] to maximize the
optimization problem (\ref{eq:FisherCrit}) under $\ell_{1}$-penalization.

\begin{figure}

\includegraphics{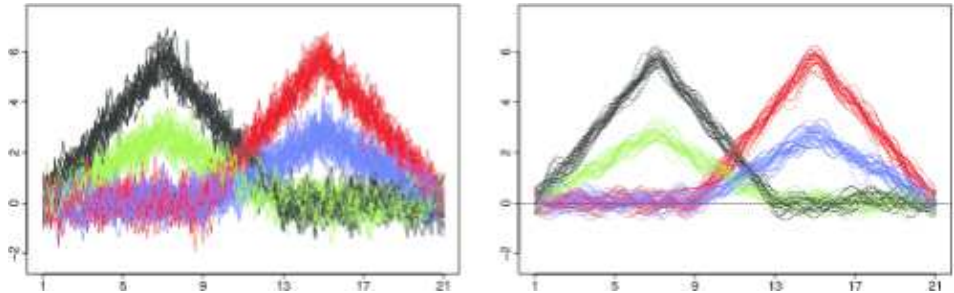}

\caption{Raw and smoothed simulated curves.} \label{fig:SimulatedCurves}
\end{figure}

\section{Numerical experimentations}\label{sec4}

This section presents numerical experiments to validate on simulated
and benchmark data the approach presented above, before applying it on
the BSS data.
\subsection{Model selection}\label{sec4.1}

We first focus on the problem of model selection. Here, BIC and the
slope heuristic are challenged on a set of simulated curves. A sample
of $n=100$ curves is simulated according to the following model,
inspired by \citet{Fer2003,Pre2007a}:
\begin{eqnarray*}
&&\mbox{Cluster 1:}\quad X(t)  =  U+(1-U)h_{1}(t)+\epsilon(t),\qquad t\in[1,21],
\\
&&\mbox{Cluster 2:}\quad X(t) = U+(1-U)h_{2}(t)+\epsilon(t),\qquad t\in[1,21],
\\
&&\mbox{Cluster 3:}\quad X(t) = U+(0.5-U)h_{1}(t)+\epsilon (t),\qquad t
\in[1,21],
\\
&&\mbox{Cluster 4:}\quad X(t) = U+(0.5-U)h_{1}(t)+\epsilon (t),\qquad t
\in[1,21],
\end{eqnarray*}
where $U$ is uniformly distributed on $[0,1]$, and $\epsilon(t)$ is
white noise that is independent from $U$ such that $\operatorname{\mathbb{V}ar}(\epsilon_{t})=0.5$. The functions $h_{1}$ and $h_{2}$ are
defined, for $t\in[1,21]$, by $h_{1}(t)=6-|t-7|$ and
$h_{2}(t)=6-|t-15|$. The mixing proportions are equal, and the curves
are observed in $101$ equidistant points ($t=1,1.2,\ldots,21$). The
functional form of the data is reconstructed using a Fourier basis
smoothing with 25 basis functions. Figure~\ref{fig:SimulatedCurves}
plots the simulated curves and the smoothed ones.

\begin{figure}

\includegraphics{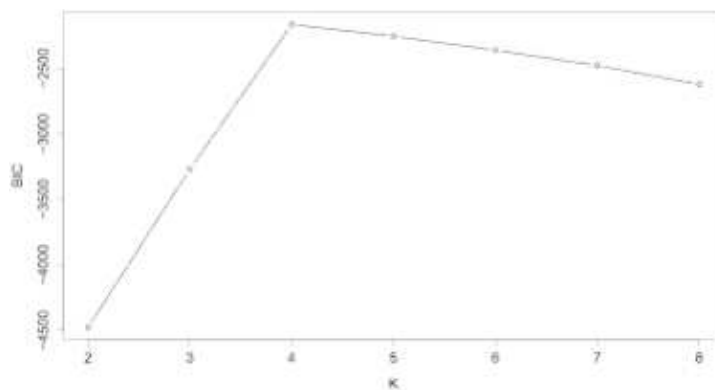}

\caption{Selection of the number of clusters
using BIC on the simulated data (actual value of $K$ is~4).} \label{fig:simul-BIC}
\end{figure}

\begin{figure}[b]

\includegraphics{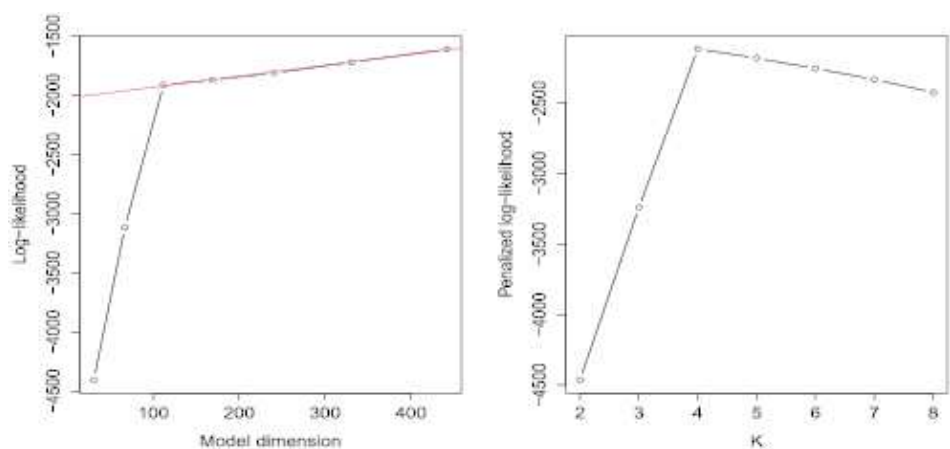}

\caption{Selection of the number of clusters
using the slope heuristic on the simulated data (actual value of $K$ is 4).}\label{fig:simul-Slope}
\end{figure}

For each simulated data set, the number $K$ of clusters is estimated
based on both the BIC and the slope heuristic criteria. As an example
of the results, Figures \ref{fig:simul-BIC} and \ref{fig:simul-Slope}
(right panel) plot, respectively, the values of the BIC criterion and
the slope heuristic for one simulation with the model $\mathrm
{DFM}_{[\Sigma_{k}\beta_{k}]}$.
On this run, both criteria succeed in selecting the actual number of
clusters ($K=4$). Figure~\ref{fig:simul-Slope} may require further
explanation. The left panel plots the log-likelihood function with
regard to the number of free model parameters, the latter being a
function of $K$ (see Table~\ref{Tab:models}). The slope heuristic
consists of using the slope of the linear part of the objective
function to calibrate the penalty. The linear part is here represented
by the red dashed line and was automatically determined using a robust
linear regression. The slope coefficient is then used to compute the
penalized log-likelihood function, shown on the right panel. We can see
here that the slope heuristic provides a penalty close to the one of BIC.

Both criteria were then used to select the appropriate model and number
of groups on 100 simulated data sets.
Tables \ref{tab:simul-result} and \ref{tab:simul-result-1} present respectively the selected
number of clusters by BIC and the slope heuristic
over 100 simulations for each of the 12 DFM models. It turns out that
although BIC can be very efficient when the model is appropriate, it
can provide unsatisfactory results in more difficult inference
situations. Conversely, the slope heuristic appears to be more
consistent in the selection of the number of clusters while keeping
very good overall results. For this reason, the selection of models and
the number of groups will be addressed in the following section with
the slope heuristic.

\begin{table}
\caption{Number of clusters selected by BIC
over 100
simulations for the 12 DFM models. Actual value for $K$ is 4}
\label{tab:simul-result}
\begin{tabular*}{\textwidth}{@{\extracolsep{\fill}}lccccccccc@{}}
\hline
& \multicolumn{9}{c@{}}{\textbf{Number} $\bolds{K}$ \textbf{of clusters}}\\[-6pt]
& \multicolumn{9}{c@{}}{\hrulefill}\\
\textbf{Model} & \textbf{2} & \textbf{3} & \textbf{4} & \textbf{5} &
\textbf{6} & \textbf{7} & \textbf{8} & \textbf{9} & \textbf{10}\\
\hline
$\mathrm{DFM}_{[\Sigma_{k}\beta_{k}]}$ & 0 & 0 & \phantom{0}\textbf{99} &
 \phantom{0}0 &
\phantom{0}0 & \phantom{0}0 & \phantom{0}0 & \phantom{0}1 & \phantom{0}0 \\
$\mathrm{DFM}_{[\Sigma_{k}\beta]}$ & 0 & 0 & \phantom{0}27 & \textbf{37} &
 23
& 12 & \phantom{0}1 & \phantom{0}0 & \phantom{0}0 \\
$\mathrm{DFM}_{[\Sigma\beta_{k}]}$ & 0 & 0 & \textbf{100} & \phantom{0}0 &
\phantom{0}0 &
\phantom{0}0 & \phantom{0}0 & \phantom{0}0 & \phantom{0}0 \\
$\mathrm{DFM}_{[\Sigma\beta]}$ & 0 & 0 & \phantom{00}2 & \phantom{0}2 & \phantom{0}8 & 10 & 10 & 10 &
\textbf{58}\\
$\mathrm{DFM}_{[\alpha_{kj}\beta_{k}]}$ & 0 & 0 & \textbf{100} & \phantom{0}0
& \phantom{0}0 & \phantom{0}0 & \phantom{0}0 & \phantom{0}0 & \phantom{0}0 \\
$\mathrm{DFM}_{[\alpha_{kj}\beta]}$ & 0 & 0 & \phantom{00}1 & \phantom{0}5 &
\phantom{0}8 & 12 & 10 &
\phantom{0}7 & \textbf{57}\\
$\mathrm{DFM}_{[\alpha_{k}\beta_{k}]}$ & 0 & 0 & \textbf{100} & \phantom{0}0 &
\phantom{0}0 & \phantom{0}0 & \phantom{0}0 & \phantom{0}0 & \phantom{0}0 \\
$\mathrm{DFM}_{[\alpha_{k}\beta]}$ & 0 & 0 & \phantom{00}0 & \phantom{0}0 &
\phantom{0}1 & \phantom{0}1 & \phantom{0}4 & \phantom{0}7 &
\textbf{87}\\
$\mathrm{DFM}_{[\alpha_{j}\beta_{k}]}$ & 0 & 0 & \textbf{100} & \phantom{0}0
& \phantom{0}0 & \phantom{0}0 & \phantom{0}0 & \phantom{0}0 & \phantom{0}0 \\
$\mathrm{DFM}_{[\alpha_{j}\beta]}$ & 0 & 0 & \phantom{0}\textbf{91} &
 \phantom{0}5 & \phantom{0}1 &
\phantom{0}1 & \phantom{0}1 & \phantom{0}0 & \phantom{0}1 \\
$\mathrm{DFM}_{[\alpha\beta_{k}]}$ & 0 & 0 & \textbf{100} & \phantom{0}0 & \phantom{0}0 &
\phantom{0}0 & \phantom{0}0 & \phantom{0}0 & \phantom{0}0 \\
$\mathrm{DFM}_{[\alpha\beta]}$ & 0 & 0 & \phantom{0}\textbf{97} &
\phantom{0}2 & \phantom{0}1 & \phantom{0}0 &
\phantom{0}0 & \phantom{0}0 & \phantom{0}0 \\
\hline
\end{tabular*}
\end{table}

\begin{table}[b]
\caption{Number of clusters selected by the slope
heuristic over 100 simulations for the 12 DFM models. Actual value for
$K$ is 4}
\label{tab:simul-result-1}
\begin{tabular*}{\textwidth}{@{\extracolsep{\fill}}lccccccccc@{}}
\hline
& \multicolumn{9}{c@{}}{\textbf{Number} $\bolds{K}$ \textbf{of clusters}}\\[-6pt]
& \multicolumn{9}{c@{}}{\hrulefill}\\
\multicolumn{1}{@{}l}{\textbf{Model}} & \multicolumn{1}{c}{\textbf{2}} & \multicolumn{1}{c}{\textbf{3}} &
\multicolumn{1}{c}{\textbf{4}} & \multicolumn{1}{c}{\textbf{5}} & \multicolumn{1}{c}{\textbf{6}} &
\multicolumn{1}{c}{\textbf{7}} & \multicolumn{1}{c}{\textbf{8}} & \multicolumn{1}{c}{\textbf{9}} &
\multicolumn{1}{c@{}}{\textbf{10}}\\
\hline
$\mathrm{DFM}_{[\Sigma_{k}\beta_{k}]}$ & \phantom{0}6 & 9 & \textbf{84} &
\phantom{0}0 &
\phantom{0}0 & 0 & 0 & 1 & 0 \\
$\mathrm{DFM}_{[\Sigma_{k}\beta]}$ & 15 & 1 & \textbf{81} & \phantom{0}3 & \phantom{0}0
& 0 & 0 & 0 & 0 \\
$\mathrm{DFM}_{[\Sigma\beta_{k}]}$ & \phantom{0}0 & 0 & \textbf{91} &\phantom{0}8 & \phantom{0}1 &
0 & 0 & 0 & 0 \\
$\mathrm{DFM}_{[\Sigma\beta]}$ & \phantom{0}0 & 0 & \textbf{77} & 17 & \phantom{0}5 & 1
& 0 & 0 & 0 \\
$\mathrm{DFM}_{[\alpha_{kj}\beta_{k}]}$ & \phantom{0}0 & 0 & \textbf{97} & \phantom{0}3
& \phantom{0}0 & 0 & 0 & 0 & 0 \\
$\mathrm{DFM}_{[\alpha_{kj}\beta]}$ & \phantom{0}0 & 0 & \textbf{65} & 17 &
14 & 3 & 1 & 0 & 0 \\
$\mathrm{DFM}_{[\alpha_{k}\beta_{k}]}$ & \phantom{0}0 & 0 & \textbf{85} & 14
& \phantom{0}1 & 0 & 0 & 0 & 0 \\
$\mathrm{DFM}_{[\alpha_{k}\beta]}$ & \phantom{0}0 & 0 & \textbf{78} & 14 & \phantom{0}7
& 1 & 0 & 0 & 0 \\
$\mathrm{DFM}_{[\alpha_{j}\beta_{k}]}$ & \phantom{0}0 & 1 & \textbf{87} & 11
& \phantom{0}1 & 0 & 0 & 0 & 0 \\
$\mathrm{DFM}_{[\alpha_{j}\beta]}$ & \phantom{0}0 & 0 & \textbf{67} &
\phantom{0}8 & \phantom{0}6 &
6 & 4 & 3 & 6 \\
$\mathrm{DFM}_{[\alpha\beta_{k}]}$ & \phantom{0}4 & 0 & \textbf{96} &
\phantom{0}0 & \phantom{0}0 &
0 & 0 & 0 & 0 \\
$\mathrm{DFM}_{[\alpha\beta]}$ & \phantom{0}0 & 0 & \textbf{87} &
\phantom{0}6 & \phantom{0}4 & 2 &
1 & 0 & 0 \\
\hline
\end{tabular*}
\end{table}

%
\begin{figure}

\includegraphics{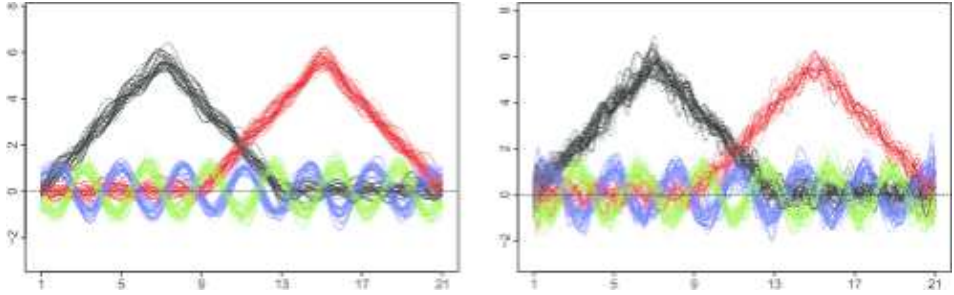}

\caption{Simulated curves with cubic spline
smoothing (left) and Fourier basis smoothing (right).} \label{fig:simul-curves2}
\end{figure}

\subsection{Selection of discriminative basis functions}\label{sec4.2}
This experiment is concerned with the selection of the discriminative
basis functions, that is, the most relevant ones for
discriminating the clusters. In this work, the selection of the
discriminative basis functions is viewed as solving the optimization
problem (\ref{eq:FisherCrit}) of the F step under sparsity constraints
(i.e., such that the loading matrix $U$ contains as many zeros
as possible). To evaluate the ability of our approach to select the
relevant discriminative basis functions, we consider now a simulation
setting in which two primarily different frequencies are involved. The
simulation setup is as follows:
\begin{eqnarray*}
&&\mbox{Cluster 1:}\quad X(t)  =  U+(1-U)h_{1}(t)+\epsilon(t),\qquad t\in[1,21],
\\
&&\mbox{Cluster 2:}\quad X(t)  =  U+(1-U)h_{2}(t)+\epsilon(t),\qquad t\in[1,21],
\\
&&\mbox{Cluster 3:}\quad X(t)  =  U+(1-U)\cos(2t)+\epsilon (t),\qquad t\in[1,21],
\\
&&\mbox{Cluster 4:}\quad X(t)  =  U+(1-U)\sin(2t-2)+\epsilon (t),\qquad t\in[1,21],
\end{eqnarray*}
where $U$, $\epsilon(t)$, $h_{1}$, $h_{2}$, the mixing proportions
where $U$, $\epsilon(t)$, $h_{1}$, $h_{2}$, the mixing proportions and
the observation points are the same as in the previous simulation
setting. The functional form of the data is reconstructed using both
Fourier basis smoothing (with 25 basis functions) and a cubic spline
basis (with 50 basis functions). Figure~\ref{fig:simul-curves2} plots
the simulated curves, respectively smoothed on cubic splines and
Fourier basis functions.
Starting from the partition estimated with FunFEM and the $\mathrm
{DFM}_{[\Sigma_{k}\beta_{k}]}$ model, the sparse version of the
algorithm is launched with the sparsity parameter $\lambda=0.1$ on
both Fourier and spline smoothed curves.

\begin{figure}

\includegraphics{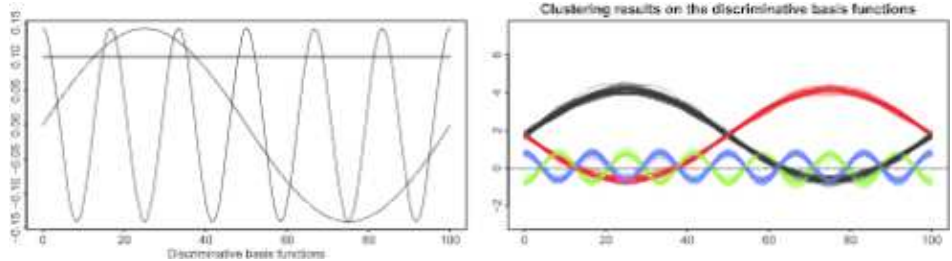}

\caption{Discriminative functions among the
Fourier basis functions: selected basis functions (left) and data
projected on the selected basis functions (right).} \label{fig:simul-curves3}
\end{figure}

\begin{figure}[b]

\includegraphics{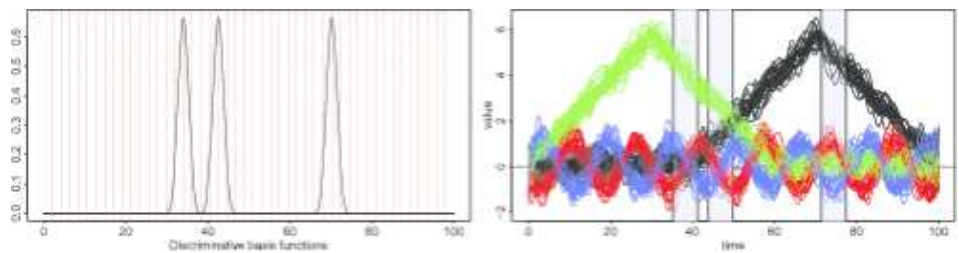}

\caption{Discriminative functions among
the spline basis functions: selected basis functions (left) and
original data with, highlighted in grey, the time periods associated
with the selected basis functions (right).}\label{fig:simul-curves3-1}
\end{figure}

Figures \ref{fig:simul-curves3} and \ref{fig:simul-curves3-1} plot
the selected basis functions on both spline and Fourier bases. For the
Fourier basis, the selection of the basis functions indicates which
periodicity in the observed curves are the most discriminative,
whereas, for the spline smoothing, it indicates which time intervals
are the most discriminant.
On the one hand, for the Fourier basis, the sparse version of FunFEM
selects only two discriminative periodicities over the $25$ original
basis functions (left panel of Figure~\ref{fig:simul-curves3}). The
selected basis functions turn out to be relevant because they actually
correspond to the two periodicities present in the simulated data. The
right panel of the figure plots the smoothed curves on the two selected
basis functions. One can observe that the basis selection is actually
relevant because the main features of the data are kept.

On the other hand, for the spline basis, sparse FunFEM has selected
three basis functions among the 25 original ones (left panel of Figure~\ref{fig:simul-curves3-1}). The three selected functions indicate the
most discriminative time intervals. Those time intervals are reported
on the right panel of the figure in addition to the curves. One can,
for instance, note that the first (from the left) selected function
discriminates the green clusters from the three other groups.
Similarly, the second discriminative function allows for separating the
black and green clusters from the blue and red curves. Finally, the
last selected function aims at discriminating the black group from the others.

\begin{table}
\caption{Clustering accuracies (in percentage) on the Kneading,
Face, ECG and Wafer data sets for FunFEM and state-of-the-art methods.
Bold results correspond to best clustering accuracies and the stars
indicate the DFM model selected by BIC}
\label{Table_3}
\begin{tabular*}{\textwidth}{@{\extracolsep{\fill}}lcccc@{}}
\hline
\multicolumn{1}{@{}l}{\textbf{Method}} & \multicolumn{1}{c}{\textbf{Kneading}} & \multicolumn{1}{c}{\textbf{ECG}} & \multicolumn{1}{c}{\textbf{Face}} &
\multicolumn{1}{c@{}}{\textbf{Wafer}}\\
\hline
kmeans-$d_{0}$ & 62.61 & 74.50 & 48.21 & 63.34\\
kmeans-$d_{1}$ & 64.35 & 61.50 & 34.80 & 62.53\\
Funclust & 66.96 & \textbf{84.00} & 33.03 & 63.10\\
FunHDDC & 62.61 & 75.00 & 57.14 & 63.41\\
Fclust & 64.00 & 74.50 & -- & --\\
Curvclust & 65.21 & 74.50 & 58.92 & 63.30\\[3pt]
FunFEM $\mathrm{DFM}_{[\Sigma_{k}\beta_{k}]}$ & 67.74 & 71.00 &
59.82 & \textbf{66.89}\\
FunFEM $\mathrm{DFM}_{[\Sigma_{k}\beta]}$ & \textbf{70.97} & 73.00
& 54.46 & 64.10\\
FunFEM $\mathrm{DFM}_{[\Sigma\beta_{k}]}$ & 67.74 & 72.00 & \textbf
{61.60} & 66.35\\
FunFEM $\mathrm{DFM}_{[\Sigma\beta]}$ & 66.66 & 75.00 & 54.46 &
64.17\\
FunFEM $\mathrm{DFM}_{[\alpha_{kj}\beta_{k}]}$ & 67.74 &  \phantom{*}71.00{*} &
 \phantom{*}53.57{*} & \textbf{66.89}\\
FunFEM $\mathrm{DFM}_{[\alpha_{kj}\beta]}$ & \textbf{70.97} & 73.50
& 54.46 & 64.10\\
FunFEM $\mathrm{DFM}_{[\alpha_{k}\beta_{k}]}$ & 67.74 & 71.00 &
53.57 & \phantom{*}\textbf{66.89{*}}\\
FunFEM $\mathrm{DFM}_{[\alpha_{k}\beta]}$ & \textbf{70.97} & 73.00
& 57.14 & 64.10\\
FunFEM $\mathrm{DFM}_{[\alpha_{j}\beta_{k}]}$ & 67.74 & 72.00 &
55.35 & 66.40\\
FunFEM $\mathrm{DFM}_{[\alpha_{j}\beta]}$ & 66.66 & 75.00 & 53.57 &
64.17\\
FunFEM $\mathrm{DFM}_{[\alpha\beta_{k}]}$ &  \phantom{*}67.74{*} & 72.00 &
53.57 & 66.40\\
FunFEM $\mathrm{DFM}_{[\alpha\beta]}$ & 66.66 & 75.00 & 56.25 &
64.17\\
\hline
\end{tabular*}
\end{table}
%

\subsection{Comparison with state-of-the-art methods}\label{sec4.3}

This last numerical study aims at comparing the FunFEM algorithm with
state-of-the-art methods on four real data sets that are commonly used
in the functional clustering literature. The data sets considered here
are as follows: the \textit{Kneading}, \textit{ECG}, \textit{Face},
and \textit{Wafer} data sets. Appendix \ref{App-A} provides a
detailed description of those data sets.

FunFEM is here compared with the six state-of-the-art methods:
kmeans-$d_{0}$ and kmeans-$d_{1}$ [\citet{Iev2012}], funclust [\citet
{Jac2013}], funHDDC [\citet{Bou2011}], fclust [\citet{Jam2003}] and
curvclust [\citet{Gia2012}]. The two kmeans-based methods use,
respectively, the $L_2$-metric between curves (kmeans-$d_0$) and
between their derivatives (kmeans-$d_1$). The four other methods assume
a probabilistic modeling. Funclust assumes a Gaussian distribution for
the functional principal components scores, whereas funHDDC, fclust and
curvclust directly model the basis expansion coefficients.

Table~\ref{Table_3} presents the clustering accuracies (according to
the known labels) on the four data sets for FunFEM and the six
clustering methods. FunFEM turns out to be very competitive with its
challengers on those data sets. FunFEM outperforms the other methods on
all data sets except the second one where it is the second best method.
On the kneading, ECG and wafer sets, the improvement over
state-of-the-art methods is significant. It is also worth noticing that
the model selected by BIC (the model associated with the higher BIC
value) often provides some of the best possible results.

\section{Analysis of bike sharing systems}\label{sec5}

This section now presents the results of the application of FunFEM to
one month of stock data from eight bike sharing systems (Managed by
JCDecaux Ciclocity and Serco) in Europe. As explained in the
\hyperref[sec1]{Introduction}, clustering is a principal way to summarize the behavior
of BSS stations, and this approach has already been used in the
literature. This study proposes going further here. The FunFEM
algorithm presents a few advantages compared to existing works for
dealing with the BSS data considered here and for comparing the eight
studied systems. First, conversely to previous works, FunFEM explicitly
addresses the functional nature of BSS stock data and, as we saw
earlier, it outperforms multivariate and functional clustering
techniques in most situations. FunFEM is therefore expected to perform
well on the BSS data and to provide meaningful clusters from the
operational point of view. Second, FunFEM is able to easily handle
large data sets, in term of time points, due to its parsimonious
modeling. This is an important point here because we consider time
series over one month (1448 time points, cf. Section~\ref{sec2}). Last
but not least, FunFEM helps visualize the clustered data into a
discriminative subspace. As we will see, this specific feature will be
particularly informative when analyzing the clustering results on the
BSS data. The visualization of the different cities within the
discriminative subspace will allow us to identify the systems with
operating issues and to propose practical solutions to improve those systems.

%

\subsection{Clustering results for Paris stations}\label{sec5.1}

\begin{figure}

\includegraphics{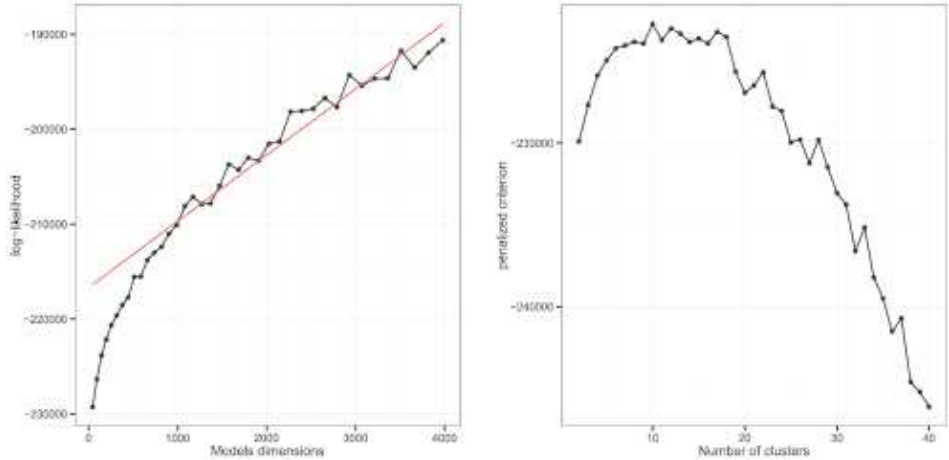}

\caption{Model selection plots for Paris:
log-likelihood with respect to model dimensionality and its estimated
linear part (left), slope heuristic criterion with respect to $K$ (right).} \label{fig:llparis}
\end{figure}

\begin{figure}

\includegraphics{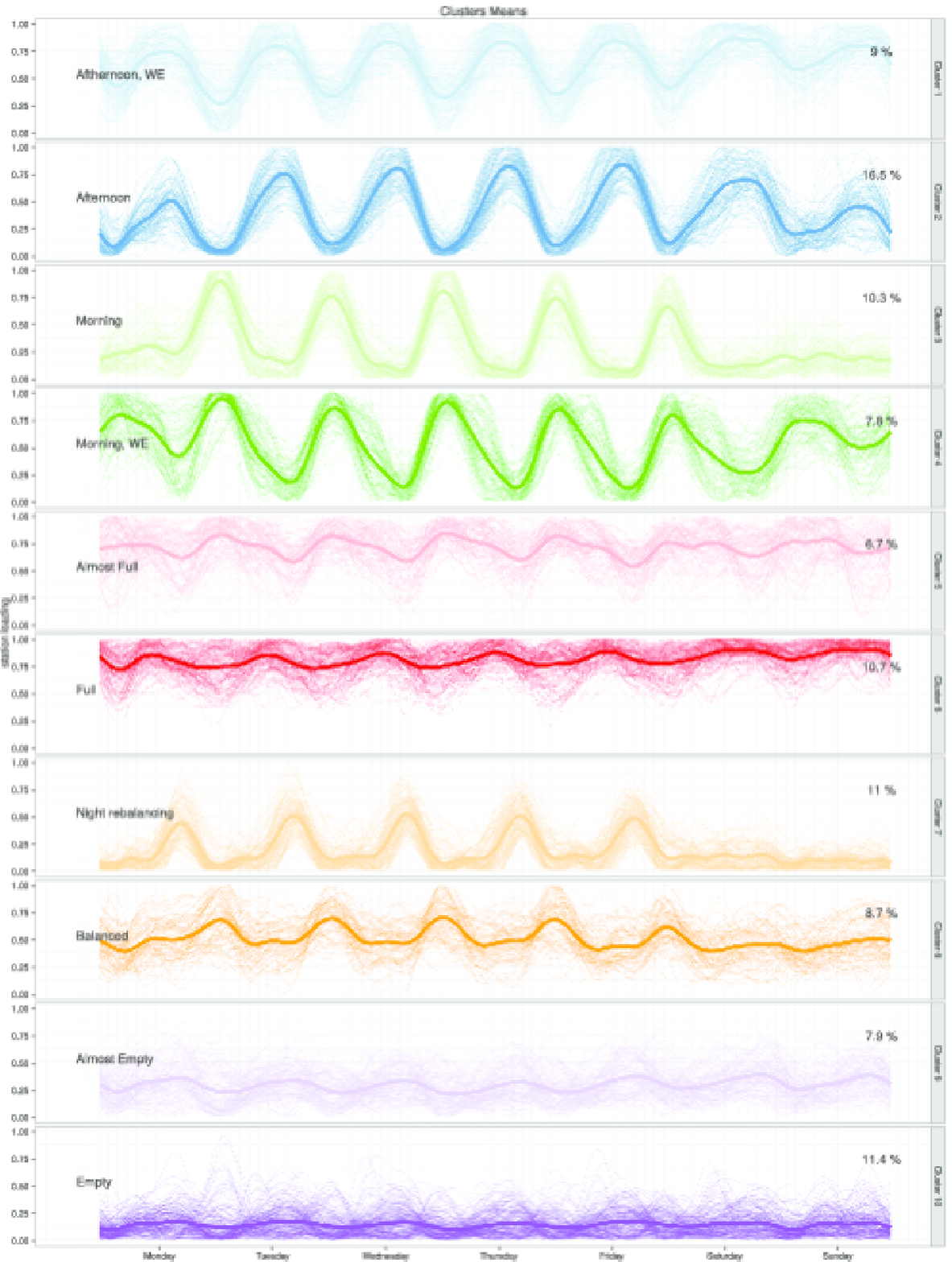}

\caption{Cluster mean profiles together
with 1000 randomly sampled curves. The name of the clusters and their
proportions are also provided.}  \label{fig:clusterprotoparis}
\end{figure}

We first begin the data analysis with solely the Paris stations. The
FunFEM algorithm has been applied on the data with a varying number of
clusters, from $2$ to $40$, and using the $\mathrm{DFM}_{[\alpha
_{kj}\beta]}$ model. This model was selected based on the good results
it obtained in the simulation study we performed. Note that it would
also be possible to test all models and select the most appropriate one
for the data using model selection. We, however, use BIC, AIC and slope
heuristic criteria to choose an appropriate value for the number $K$ of
clusters. BIC and AIC provided hard-to-use values for $K$ because even
for 40 clusters, they do not reach a maximum. Conversely, the slope
heuristic gave a satisfying value for $K$ because it reaches its
maximum for $K=10$. Figure~\ref{fig:llparis} shows the evolution of
the log-likelihood with respect to the model dimensionality and the
associated slope heuristic criterion. On the right panel, the slope
heuristic criterion peaks at $K=10$, which corresponds to an elbow in
the log-likelihood function: Above this value, the gain in
log-likelihood is linear with respect to the model dimensionality.
This value of $K$ was used for the cluster analysis. The mean profiles
of the obtained clusters are depicted in Figure~\ref{fig:clusterprotoparis}, together with the cluster proportions and a
sample of curves that belong to each cluster.

The obtained clusters are fairly balanced, with approximately ten
percent of the stations in each. The clusters are also easily
distinguishable. The stations of the first two clusters get bikes
during the afternoon and the evening. These stations differ during the
weekend; the first cluster presents high values throughout this period,
whereas the second cluster experiences a lack of bikes on Saturday mornings.
Taking into account these observations, we named the first cluster
\textit{Afternoon}, \textit{Weekend} and the second \textit{Afternoon} as a
reference to the periods where these stations are full. The next two
clusters present a phase opposition with respect to the previous ones;
these stations are full at the end of the morning rush hour
(approximately 9 a.m.). Because these two clusters differ in their
weekend behavior, we named the first one \textit{Morning} because
these stations are almost empty throughout the entire weekend, and we
named the second one \textit{Morning}, \textit{Weekend} because bikes are
available at these stations for a good part of the weekend. The next
two clusters do not present the same types of variations; their loading
profiles are considerably stable throughout the week. The difference is
in the level of fullness, with one cluster loading at approximately
0.85 and one at approximately 0.7. The first cluster also presents day
variations that are not visible for the second one. We named these
clusters \textit{Full} and \textit{Almost Full}.

\begin{figure}

\includegraphics{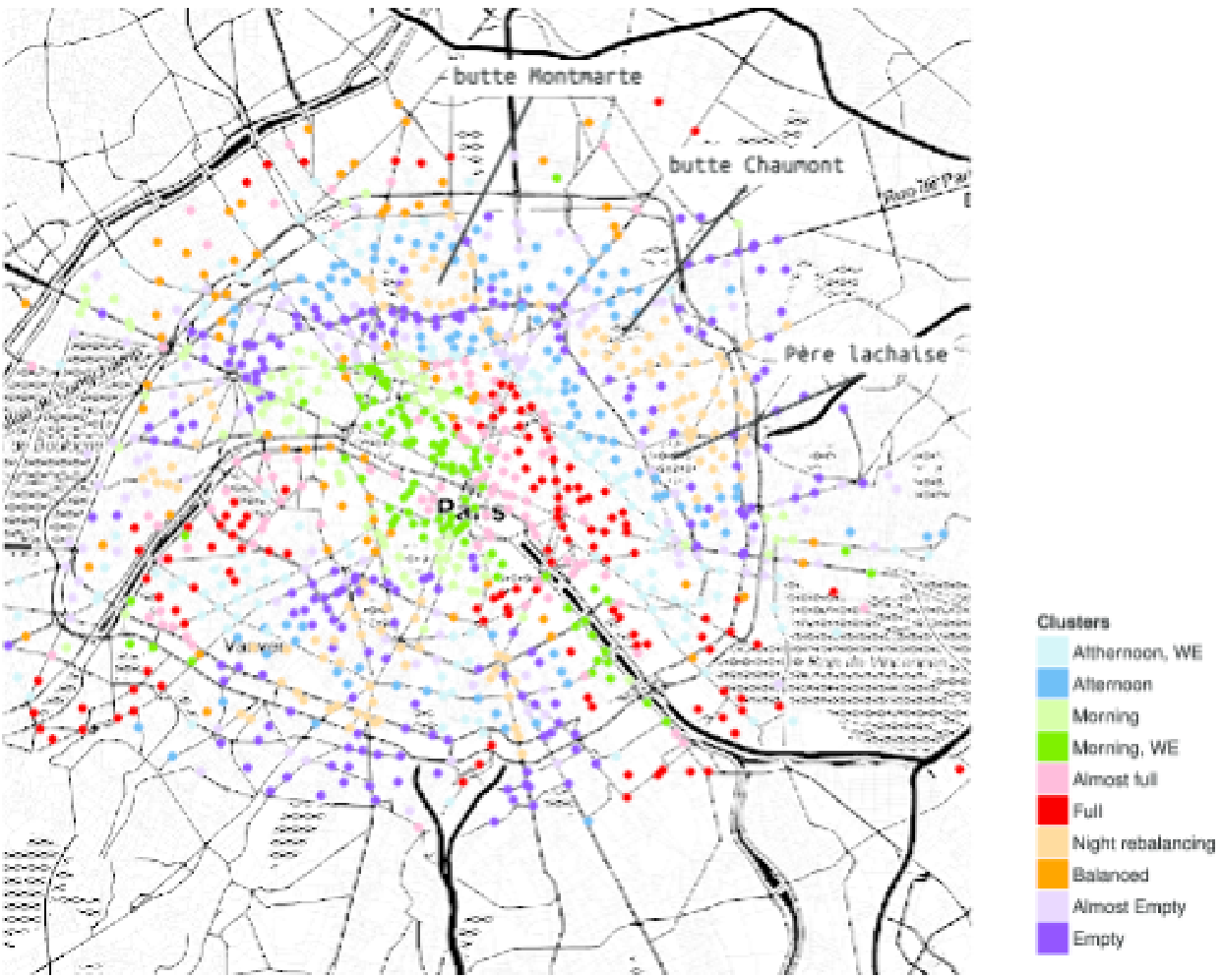}

\caption{Map of the clustering results for Paris
stations.}  \label{fig:mapparis}
\end{figure}

Clusters 7 and 8 present overall small activity: Cluster 7 get bikes at
night, but this does not saturate the stations that reach a balanced
state in these time periods. This phenomenon may be due to the
reallocation journey performed by the operator to balance the system at
night. Cluster 8 oscillates around a balanced state, receiving slightly
more bikes during the afternoons. Taking into account these remarks, we
call these clusters \textit{Night rebalancing} and \textit{Balanced}.
Finally, clusters 9 and 10 gather stations that are almost empty
throughout the week. Cluster 9 presents considerably stable behavior
with a constant loading profile of approximately 0.25, whereas the
second one smoothly oscillates at approximately 0.1. We respectively
call these clusters \textit{Almost empty} and \textit{Empty}.

To complement this analysis of the clustering results, Figure~\ref{fig:mapparis}\footnote{Map build using the ggmap package for R [\citet
{Kahle2013}].} presents the spatial location of the clustering results.
One of the first things that catches the eye when looking at this
figure concerns the relatively good spatial organization of the
results, although this information was never used in the clustering
process. Stations from the same clusters are frequently grouped
together on the map. From a Parisian perspective, those results are
natural: The \textit{Morning} and \textit{Morning}, \textit{Weekend} clusters
(in green on the map) are located in areas with a high employment
density, which therefore correspond to destinations during the morning
commute. This phenomenon explains why these stations experience a
saturation at the end of the morning rush hour. On the contrary, the
blue clusters, which correspond to the \textit{Afternoon} and \textit
{Afternoon}, \textit{Weekend} clusters, are located in more residential
neighborhoods with a higher population density. They therefore
correspond to classical origins during the morning rush hour and lose
their bikes during this time period. The stations that belong to these
clusters are located in regions that are close to \textit{Empty},
\textit{Almost empty} stations, which are more problematic from a user
perspective. These neighborhoods are not in the hyper-center of Paris,
and they are also located close to stations that belong to the \textit
{Night rebalancing} cluster. The \textit{Night rebalancing} cluster is
frequently located in uphill locations, such as the ``Butte Montmarte,''
the ``P\`ere Lachaise'' cemetery and the ``Butte Chaumont'' garden.
Finally, the \textit{Full} and \textit{Almost full} stations are
located in the center, whereas the \textit{Balanced} stations are
located primarily in the periphery of the system.

In comparison with previous results obtained based on Paris bike share
origin/destination data, such as in \citet{come2014}, these
observations are considerably consistent. One of the major differences
concerns parks and leisure locations, which do not emerge from the
clustering in our study. This phenomenon may be explained by the
difference in the nature of the input data. The stock data that are
used in this paper do not enable the differentiation of these stations,
whereas origin/destination data do. However, stock data are easier to
obtain on a large scale and thus will allow cross-city comparisons,
which is the subject of the next section.

\subsection{Clustering results on several cities}\label{sec5.2}

\begin{figure}

\includegraphics{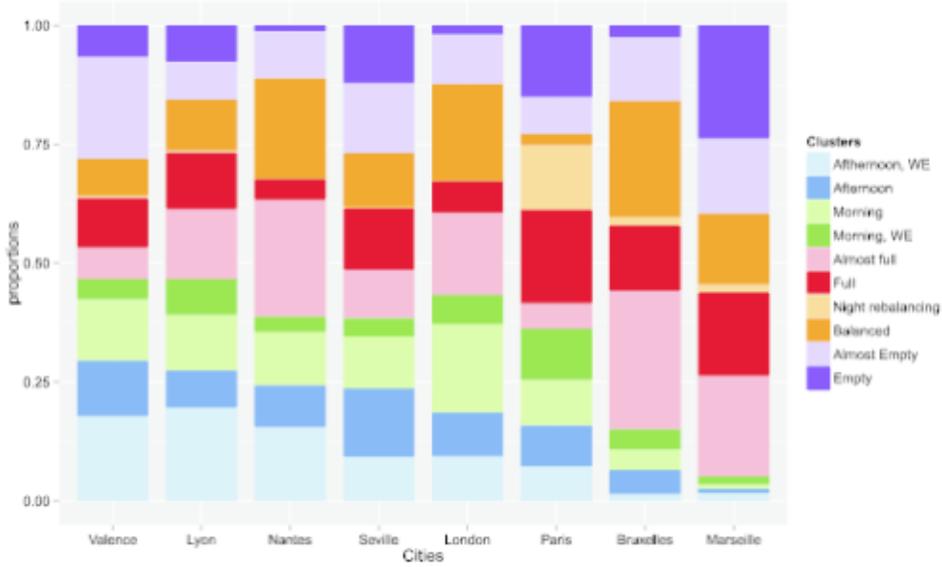}

\caption{Cluster proportions by city.} \label{fig:clusterprop}
\end{figure}

The clustering was also performed on the entire data set, which
includes stations from the eight systems (see Table~\ref{tab:bsslist}). The same methodology was used; the curves were
projected on the same Fourier basis, and, as prior, the clustering was
performed with the model $\mathrm{DFM}_{[\alpha_{kj}\beta]}$ and
with a varying number of clusters, from 2 to 40. The slope heuristic
leads to the same number of clusters ($K=10$ clusters) in this larger
data set. The obtained clusters are also close to those obtained only
in Paris. Their profiles, which are supplied in the \hyperref[app]{Appendix}, are close
to those shown in Figure~\ref{fig:clusterprotoparis}, and their
interpretation does not differ significantly. We kept the same labels
for the clusters because the main difference comes from the amplitude
of the profile variations, which are smaller in the entire data set. An
interesting point in the obtained results concerns the proportions of
the different clusters for each city. This indeed enables an aggregate
view of the systems that eases their comparison. These proportions are
shown in Figure~\ref{fig:clusterprop}.

Differences between the cities are visible in this figure. The
proportion of the \textit{Night rebalancing} cluster is, for example,
much more important for Paris than in any other city. This cluster,
which corresponds to stations that are rebalanced during the night, is
not visible in cities other than Paris. On the contrary, the proportion
of the \textit{Balanced} cluster is much smaller in Paris than in the
other cities. Another clear difference concerns the \textit{Empty} and
\textit{Almost empty} cluster stations, which are important in
Marseille and Bruxelles. In Marseille, the \textit{Full} and \textit
{Almost full} clusters are also over-represented, corresponding to more
than 25\% of the city stations. This system seems, therefore, the more
unbalanced system with many stations frequently full or empty.
Conversely, the cities on the left of the plot, such as Valencia or
Lyon, seem to be more active and balanced, with an important proportion
of stations that belong to the \textit{Afternoon} and \textit
{Morning} clusters. This aggregate view helps identify the BSSs that do
not have satisfying behavior from the exploitation point of view.
Indeed, Bruxelles and Marseille have exploitation profiles with low or
even very low proportions of the active clusters (\textit
{Afternoon}, \textit{WE}, \textit{Afternoon}, \textit{Morning},
\textit{WE} and \textit
{Morning}). Conversely, the BSS of Valencia, Lyon and London seem to be
the most efficient systems. Some of the factors that may explain these
behaviors are the ratio between bikes and docks, the topography and
geography of the cities and the bike redistribution policy.

\begin{figure}

\includegraphics{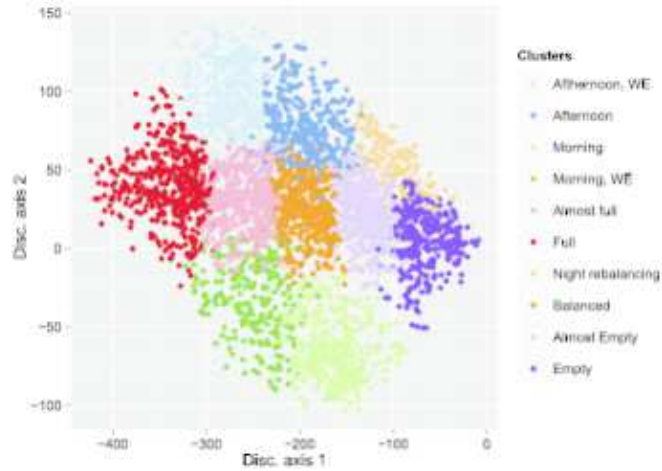}

\caption{Bike stations projected into the two
first axes of the discriminative functional subspace. Colors indicate
the cluster memberships.}  \label{fig:disc-axis}
\end{figure}

\begin{figure}[b]

\includegraphics{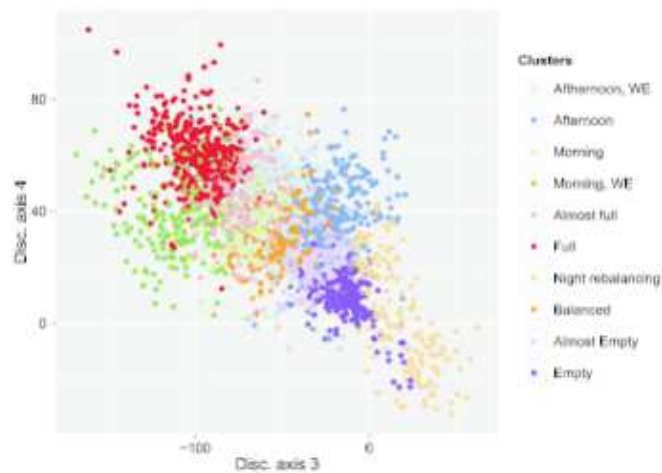}

\caption{Bike stations projected into the
third and fourth axes of the discriminative functional subspace. Colors
indicate the cluster memberships.} \label{fig:disc-axis-bis}\vspace*{6pt}
\end{figure}

\begin{figure}

\includegraphics{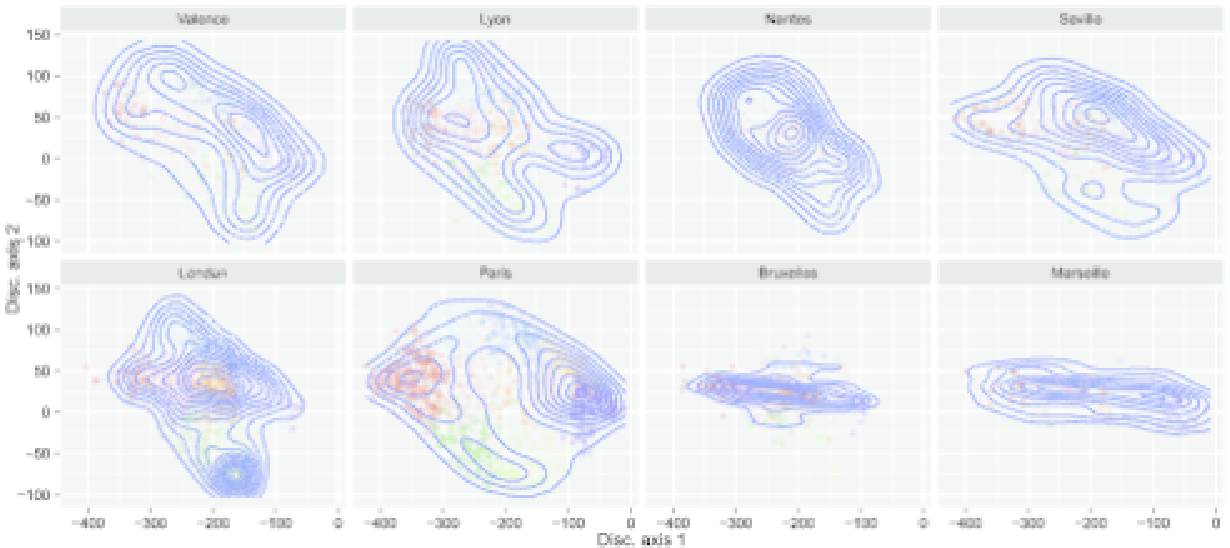}

\caption{Density of bike stations per
city projected into the two first axes of the discriminative functional
subspace.}\label{fig:disc-axis-facet}
\end{figure}

The observations made on the cluster proportions from Figure~\ref{fig:clusterprop} can be confirmed by looking at the discriminative
functional subspace estimated by FunFEM. Figure~\ref{fig:disc-axis}
shows the bike stations of the 8 cities projected into the two first
axes of the discriminative subspace. Figure~\ref{fig:disc-axis-bis}
shows the projection into the third and fourth discriminative axes. The
colors indicate the cluster memberships of the stations. It may first
be useful to interpret the discriminative axes from the cluster
meanings. The first axis puts in opposition the \textit{Full} and
\textit{Empty} clusters and can be therefore viewed as a station
loading axis. The second axis opposes the \textit{Afternoon} and
\textit{Morning} clusters. It can therefore be linked with the phase
of the curves. The third and fourth axes are less interpretable and
seem primarily linked with the \textit{Night rebalancing} cluster.
Knowing the meaning of the discriminative subspace axes enables the
comparison of the studied systems through the analysis of their station
behaviors. Figure~\ref{fig:disc-axis-facet}\ shows the projection of
the bike stations for each city on the two first axes of the
discriminative subspace. A kernel density estimation is also proposed
to visualize the relative density of stations in this subspace. This
visual representation confirms the first comparison results of the
cluster proportions. In particular, Marseille and Bruxelles present a
distribution in the discriminative subspace that is considerably
different from that of other cities. Indeed, both are oriented along
the first discriminative axis and do not present significant variations
along the second axis. The signature of those two cities within the
subspace can be qualified as problematic from an operational point of
view because the first discriminative axis opposes the \textit{Full}
and \textit{Empty} clusters, whereas the second axis is associated
with the \textit{Afternoon} and \textit{Morning} clusters.

The spatial analysis of the results was also performed by mapping the
clustering results (see Figure~\ref{fig:mapall} in
Appendix \ref{appb}). As
with Paris, it turns out that the different clusters are also
frequently spatially clustered. Furthermore, the same type of global
organization is visible for the different cities. Stations from the
\textit{Morning} clusters are located in the center of the systems,
whereas the other clusters are located in the periphery of the system.

\subsection{Recommendations for BSS operators}\label{sec5.3}

\begin{figure}

\includegraphics{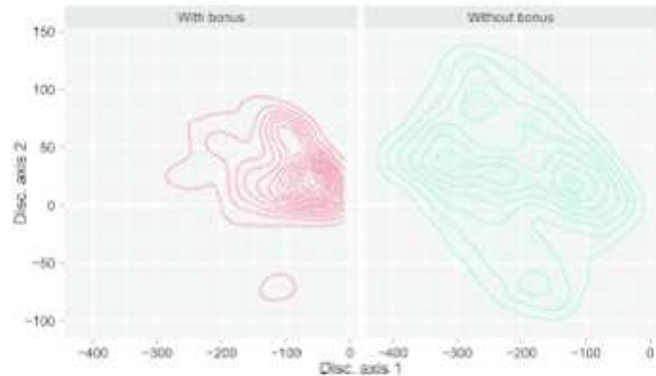}

\caption{Density of ``bonus'' (left) and regular
stations (right) projected into the two first axes of the
discriminative functional subspace.} \label{fig:Bonus}\vspace*{9pt}
\end{figure}

In light of the analyses and comparisons made above, it is possible to
make some recommendations for BSS operators regarding system structures
and policies. On the one hand, the BSS systems of Marseille and
Bruxelles appear to be composed primarily of \textit{Full} and \textit
{Empty} stations, which necessarily implies user dissatisfaction.
Possible ways to improve these situations would be either to use a
``bonus'' policy or to increase the rebalancing performed by the
operator. The ``bonus'' policy is attractive for both users and
providers. It consists of offering extra free minutes of bike usage to
users willing to return the bike to elevated stations. In theory, this
strategy should help rebalance the system. Bonus stations, for
instance, are available in Paris and Bruxelles. Thanks to our
discriminative subspace, it is in fact possible to check the real
effect of such a policy. Figure~\ref{fig:Bonus} shows the density of
``bonus'' and regular stations within the two first axes of the
discriminative functional subspace. It appears that the effect of the
bonus policy is globally limited because there is no significant
distribution difference between the regular and bonus statuses of the
\textit{Full} stations (stations projected on the left of the first
discriminative axis). We therefore recommend to BSS operators to either
modify their bonus policies (e.g., extra time bonus, cash reward) or to
increase the nighttime rebalancing of the stations for those two cities.

However, the comparison of the largest and most efficient systems has
highlighted some weaknesses of the Paris system. Although the Velib
system is one of the largest and most popular systems in the world, it
also appears to have too many \textit{Full} and \textit{Empty}\vadjust{\goodbreak}
stations, particularly compared to London. The night rebalancing
operated by JCDecaux seems to be efficient but not sufficient to
completely solve this issue. As we have seen, shared bikes are used
primarily for home-work journeys, and the stations from the \textit
{Afternoon} and \textit{Morning} clusters therefore play a key role in
the system efficiency. This situation emphasizes the importance of
commutes in the use of the service, and city bike policies must
seriously consider this aspect when designing bike paths. London's
``Cycle Superhighways'' initiative, which connects suburbs with the city
center, seems particularly effective with respect to this point in our
analyses. Those specific bike paths indeed connect stations from the
\textit{Afternoon} and \textit{Morning} clusters (see Appendix \ref{appb}). We
therefore recommend to city planners to develop bike paths in a similar
way to improve the performance of system commutes.

\section{Conclusion}\label{sec6}

This work was motivated by interest in analyzing and comparing several
European BSSs to identify common operating patterns and to propose
practical solutions to avoid potential issues. To this end, the
discriminative functional mixture (DFM) model was proposed to model the
functional data generated by the systems. In this framework, the data
are modeled into a discriminative functional subspace. The FunFEM
algorithm has been proposed for the inference of the DFM model. The
selection of the most discriminative basis functions can also be made
afterward by introducing sparsity through a $\ell_1$-type
penalization. Numerical experiments have demonstrated the efficiency of
the proposed clustering technique for both simulated and benchmark
data. FunFEM appears to be a good challenger to the best
state-of-the-art methods. The numerical experiments have also shown the
good behavior of the ``slope heuristic'' for model selection in this
context.

The proposed methodology has been applied to one-month usage statistics
of 8 bike sharing systems. FunFEM presents several advantages over
existing works for analyzing and comparing bike sharing systems. FunFEM
benefits from its parsimonious modeling and its discriminative
subspace. The obtained results were easily interpretable and useful to
obtain a compact representation of BSS system behaviors. In particular,
the discriminative subspace appears to be a useful tool to compare the
different systems with regard to the identified operating patterns.
Recommendations to BSS operators are made based on the clustering
results.

Finally, the discriminative subspace offers an interesting tool from an
operational point of view to track changes in the behavior of bike
stations. Using a sliding window and projecting the station functional
description within this window into the discriminative subspace, one
may obtain a trajectory for each station within the subspace, allowing
for the detection of any changes in the station behavior. This may be
useful when trying new pricing or bonus policies to check their effects
on the system.

\begin{appendix}\label{app}
\section{Additional information about the benchmark data sets}\label{App-A}
The Kneading data set [\citet{Lev2004}] comes from Danone VitaPole
Paris Research Center and concerns the quality of cookies and the
relationship with the flour kneading process. There are 115 different
flours for which the dough resistance is measured during the kneading
process for 480 seconds. The data set contains 115 kneading curves
observed at 241 equispaced instants of time in the interval $[0, 480]$.
The $115$ flours produce cookies of different quality: $50$ of them
produced cookies of \textit{good} quality, $25$ produced \textit
{medium} quality, and $40$ produced \textit{low} quality. Following
\citet{Lev2004,Pre2007}, least squares approximation based on cubic
B-spline functions (with 18 knots) is used to reconstruct the true
functional form of each sample curve. The ECG, Face and Wafer data sets
are benchmarks taken from the \textit{UCR Time Series Classification
and Clustering} website.\footnote{\url{http://www.cs.ucr.edu/\textasciitilde eamonn/time\_series\_data/}.} The ECG data set consists of 200
electrocardiograms from 2 groups of patients sampled at 96 time
instants and has already been studied in \citet{Ols2001}. The Face data
set [\citet{Xi2006}] consists of 112 curves sampled from 4 groups at
350 instants of time. The Wafer data set [\citet{Ols2001}] consists of
7174 curves sampled from 2 groups at 152 instants of time. For these
three data sets, the same basis of functions as for the kneading data
set has been arbitrarily chosen (20 cubic B-splines).
\newpage

\section{Detailed clustering results on the 8 BSS}\label{appb}\vspace*{-15pt}
\renewcommand{\thefigure}{B.\arabic{figure}}
\setcounter{figure}{0}
\begin{figure}[b]

\includegraphics[scale=0.99]{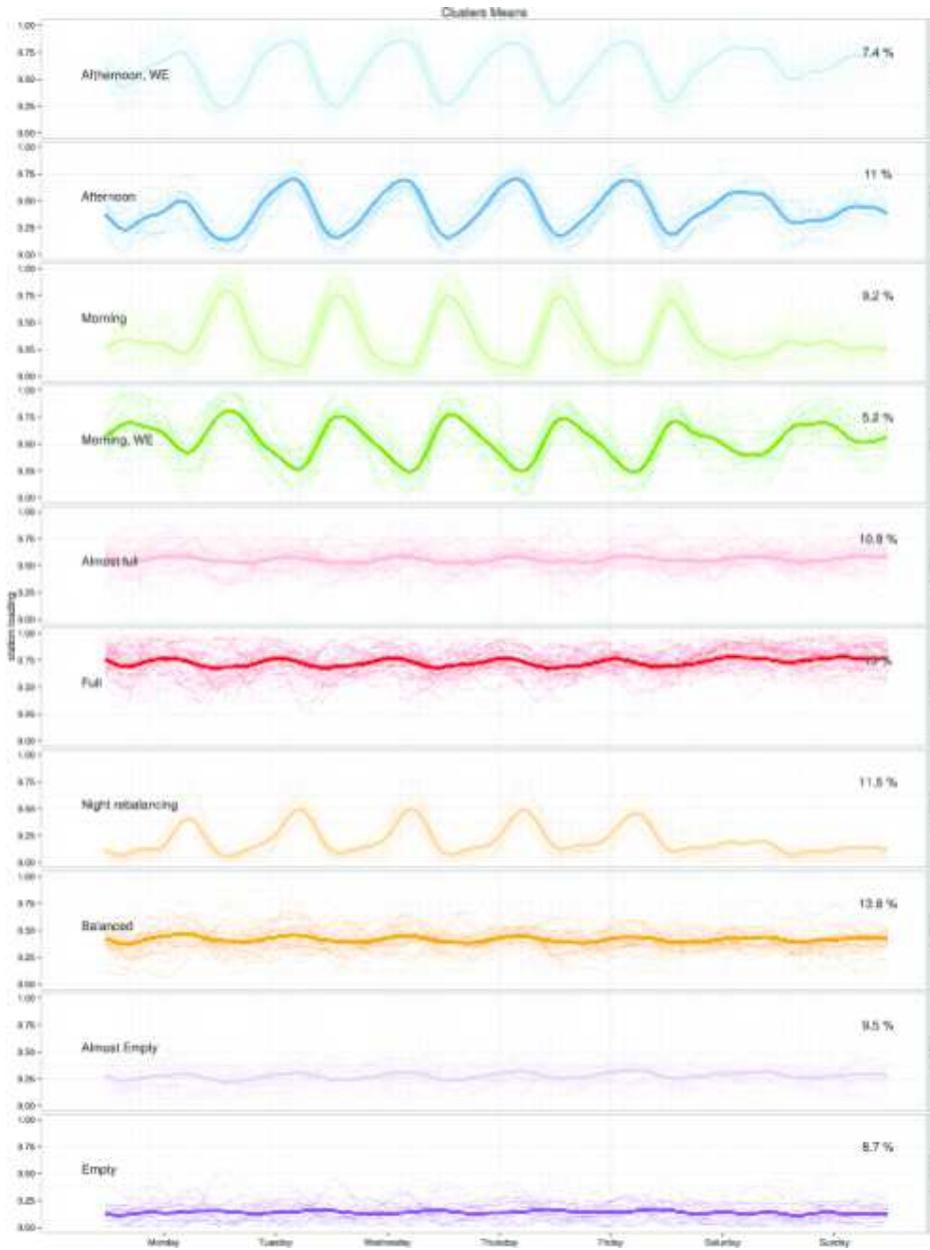}

\caption{Cluster mean profiles together with
1000 randomly sampled curves for the whole data set (Paris, London,
Bruxelles, Lyon, Valencia, Sevilla and Nantes).} \label{fig:clustall}
\end{figure}

\begin{figure}

\includegraphics{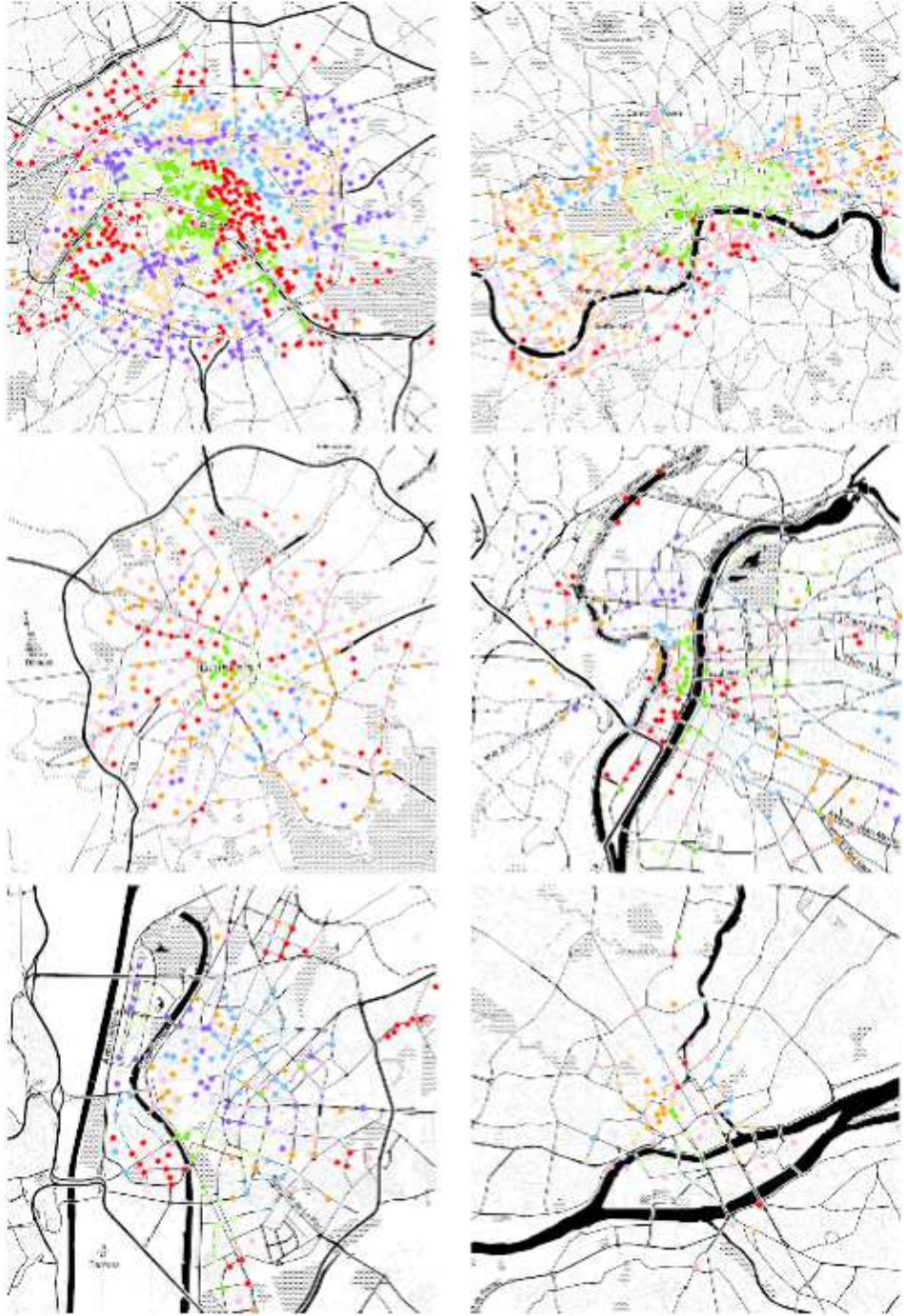}

\caption{Maps of the clustering results (from left
to right and top to bottom) for Paris, London, Bruxelles, Lyon,
Valencia, Sevilla and Nantes.} \label{fig:mapall}
\end{figure}

\end{appendix}

\mbox{}
\newpage
\section*{Acknowledgments}
The authors would like to thank the Editors and the reviewers for their
meaningful comments which have greatly contributed to improving the manuscript.

%





\printaddresses
\end{document}